\documentclass[a4paper,11pt]{article}
\pdfoutput=1 % if your are submitting a pdflatex (i.e. if you have images in pdf, png or jpg format)

\usepackage{jcappub} % for details on the use of the package
\makeatletter
\gdef\@fpheader{}
\g@addto@macro\bfseries{\boldmath}
\makeatother

\usepackage[OT1]{fontenc} % if needed

\def\baq{\begin{eqnarray}}
\def\eaq{\end{eqnarray}}

\def\beq{\begin{equation}}
\def\eeq{\end{equation}}

\usepackage{xcolor}
\usepackage[normalem]{ulem} % needed for strikethrough \sout command
\usepackage{cancel} % strikethrough \cancel command for maths
\usepackage{soul}

\usepackage{ifthen}
\usepackage{tikz}
\usepackage{physics,amsmath,bm}
\usepackage{xspace}
\usepackage[T1]{fontenc}

\usepackage{amsmath, nccmath}
\usepackage{amsfonts} 
\usepackage{blindtext}

\usepackage[utf8]{inputenc}

\usepackage{tikz-feynman}
\usepackage{amssymb}
\usepackage{graphicx}
\usepackage{physics}
\usepackage{comment}
\usepackage{slashed}

\newcounter{NumArgs}

\newcommand{\Eqs}[1]{\setcounter{NumArgs}{0}\foreach\i in{#1}{\stepcounter{NumArgs}}%
\ifthenelse{\equal{\theNumArgs}{1}}{Eq.~(\ref{#1})}%
{\ifthenelse{\equal{\theNumArgs}{2}}%
{Eqs.~\foreach\i[count=\q]in{#1}{\ifthenelse{\equal{\q}{\theNumArgs}}{and (\ref{\i})}{(\ref{\i})~}}}%
{Eqs.~\foreach\i[count=\q]in{#1}{\ifthenelse{\equal{\q}{\theNumArgs}}{and (\ref{\i})}{(\ref{\i}),~}}}}}

\newcommand{\refs}[1]{\setcounter{NumArgs}{0}\foreach\i in{#1}{\stepcounter{NumArgs}}%
\ifthenelse{\equal{\theNumArgs}{1}}{(\ref{#1})}%
{\ifthenelse{\equal{\theNumArgs}{2}}%
{\foreach\i[count=\q]in{#1}{\ifthenelse{\equal{\q}{\theNumArgs}}{and (\ref{\i})}{(\ref{\i})~}}}%
{\foreach\i[count=\q]in{#1}{\ifthenelse{\equal{\q}{\theNumArgs}}{and (\ref{\i})}{(\ref{\i}),~}}}}}

\newcommand{\Figs}[1]{\setcounter{NumArgs}{0}\foreach\i in{#1}{\stepcounter{NumArgs}}%
\ifthenelse{\equal{\theNumArgs}{1}}{Figure~\ref{#1}}%
{\ifthenelse{\equal{\theNumArgs}{2}}%
{Figures~\foreach\i[count=\q]in{#1}{\ifthenelse{\equal{\q}{\theNumArgs}}{and \ref{\i}}{\ref{\i}~}}}%
{Figures~\foreach\i[count=\q]in{#1}{\ifthenelse{\equal{\q}{\theNumArgs}}{and \ref{\i}}{\ref{\i},~}}}}}

\newcommand{\Gen}[2]{\setcounter{NumArgs}{0}\foreach\i in{#2}{\stepcounter{NumArgs}}%
	\ifthenelse{\equal{\theNumArgs}{1}}{#1.~(\ref{#2})}%
	{\ifthenelse{\equal{\theNumArgs}{2}}%
		{#1.~\foreach\i[count=\q]in{#2}{\ifthenelse{\equal{\q}{\theNumArgs}}{and (\ref{\i})}{(\ref{\i})~}}}%
		{#1.~\foreach\i[count=\q]in{#2}{\ifthenelse{\equal{\q}{\theNumArgs}}{and (\ref{\i})}{(\ref{\i}),~}}}}}

\usepackage{caption}
\usepackage{subcaption}

\usepackage{float}
\usepackage{lipsum}  
\title{Higgs-portal dark matter beyond kinetic equilibrium}
\author[d,e]{Giorgio Arcadi}
\author[a,b]{Michele Benaco}
\author[f]{Mattia Di Mauro}
\author[a,b]{Sami Nurmi}
\author[c,b]{Kimmo Tuominen}

\affiliation[a]{Department of Physics, P.O.Box 35 (YFL), FIN-40014 University of Jyväskylä, Finland}
\affiliation[b]{Helsinki Institute of Physics, PL 64, 00014 University of Helsinki, Finland}
\affiliation[c]{Department of Physics, P.O. Box 64, FIN-00014 University of Helsinki, Finland}
\affiliation[d]{Dipartimento di Scienze Matematiche e Informatiche, Scienze Fisiche e Scienze della Terra, Universita degli Studi di Messina, Via Ferdinando Stagno d'Alcontres 31, I-98166 Messina, Italy}
\affiliation[e]{INFN Sezione di Catania, Via Santa Sofia 64, I-95123 Catania, Italy}
\affiliation[f]{Istituto Nazionale di Fisica Nucleare, via P. Giuria 1, 10125 Torino, Italy}

\emailAdd{michele.m.benaco@jyu.fi}
\emailAdd{sami.t.nurmi@jyu.fi}
\emailAdd{kimmo.i.tuominen@helsinki.fi}
\emailAdd{giorgio.arcadi@unime.it}
\emailAdd{dimauro.mattia@gmail.com}
\abstract{
We revisit the scalar, vector, and fermion Higgs-portal dark matter models in the resonant regime without assuming kinetic equilibrium during the freeze-out. We compute the relic abundance by solving the full momentum-dependent Boltzmann equation and compare the results with the standard momentum-averaged treatment. 
We find that the standard treatment significantly underestimates the relic abundance below the resonance, apart from the fermionic case with a purely scalar portal coupling where the difference is negligible. The deviation from the standard treatment is largest for fermionic models with pseudoscalar interactions, corresponding to factor of 10 in terms of the coupling that reproduces the correct relic abundance. For the scalar and vector cases the corresponding difference is up to a factor of 2. We confront the resulting parameter space with the latest constraints from direct and indirect detection, as well as collider experiments.  
We find the full momentum-dependent results systematically strengthen the direct detection constraints. 

}

\begin{document}
{\hfill{\footnotesize{}}}

\maketitle
\raggedbottom
%%%%%%%%%%%%%%%%%%%%%%%%%%%%%%%%%
\section{Introduction}
\label{sec:intro}

The nature of dark matter (DM) remains one of the outstanding questions in cosmology and particle physics, and provides a strong indication of physics beyond the Standard Model (SM). 
Among the simplest and most economical extensions of the SM that provide viable DM candidates are Higgs-portal models, in which the dark sector communicates with ordinary matter through the Higgs boson (see Ref.~\cite{Arcadi:2019lka} for a review). 
These models have been studied extensively, and the interplay of relic-density, direct-detection, indirect-detection, and collider constraints excludes a large fraction of their parameter space~\cite{Cline:2013gha,Fedderke:2014wda,Beniwal:2015sdl, GAMBIT:2017gge, GAMBIT:2018eea, DiMauro:2023tho}.
%{\color{olive}{add some references here}}.
A notable exception is the resonant region,
\[
    m_{\rm DM} \simeq \frac{m_h}{2},
\]
where DM annihilates through an almost on-shell Higgs boson. 
The resonant enhancement allows the observed relic abundance to be obtained with small portal couplings, which in turn suppress signals in direct-detection and collider experiments.

The same feature that keeps the resonant region phenomenologically viable also makes the relic-density calculation theoretically delicate. 
Close to the resonance, the annihilation cross section varies rapidly with the centre-of-mass energy, making the annihilation rate sensitive not only to the DM mass and coupling but also to the momentum distribution of DM particles during freeze-out.
The standard treatment~\cite{Gondolo:1990dk} assumes that elastic scattering with the thermal bath maintains kinetic equilibrium throughout chemical decoupling,
so that the phase-space distribution retains its equilibrium shape and the Boltzmann equation reduces to a closed equation for the number density.
%The standard treatment is based on the momentum integrated Boltzmann equation ~\cite{Gondolo:1990dk}. This treatment assumes that elastic interaction with the thermal bath remain efficient enough to maintain kinetic equilibrium during the entire chemical decoupling.
%Under this assumption, the DM phase-space distribution keeps its equilibrium shape, so that the full Boltzmann equation can be reduced to a closed equation for the number density. 
Although this approximation is successful in many thermal DM scenarios, it can fail when kinetic decoupling occurs early or when the annihilation rate has a strong momentum dependence. 
%Such situations 
Capturing these effects requires a more complete treatment based on moment equations or, more generally, on the full Boltzmann equation for the phase-space distribution~\cite{Binder:2017rgn,Ala-Mattinen:2019mpa, Ala-Mattinen:2022nuj}.

The importance of early kinetic decoupling near the Higgs resonance was first demonstrated for scalar singlet DM~\cite{Binder:2017rgn}, and this case
has since been analysed using the full phase-space distribution~\cite{Ala-Mattinen:2019mpa,Ala-Mattinen:2022nuj,DiMauro:2023tho}. 
For fermionic DM with CP-conserving and CP-violating Higgs portal couplings, an analysis based on moment equations was presented in Ref.~\cite{Abe:2020obo}. 
To our knowledge, a uniform treatment of all three DM spin assignments
at the level of the full phase-space Boltzmann equation,
combined with up-to-date experimental constraints, has not yet been presented.

In this work we provide such an analysis.
%In this work, in order to cover the main phenomenological possibilities, we apply this full phase-space treatment to scalar, vector and fermionic Higgs-portal DM in the resonant region. 
We solve the full phase-space Boltzmann equation for scalar, vector, and Dirac-fermion Higgs-portal DM in the resonant region using the code \texttt{DRAKE}~\cite{Binder:2021bmg} and compare the results with the standard number-density treatment.
The scalar case provides a minimal renormalizable model, whereas the vector case is treated here in an effective Proca description. Renormalizable, gauge-invariant realizations of vector DM can be obtained when the vector field arises from a spontaneously broken dark gauge symmetry. 
%By contrast, the fermionic case is described 
The fermion couples through a dimension-five operator, which we interpret within a low-energy effective theory. 
Concrete ultraviolet completions of Higgs-portal models with fermionic and vector DM have been discussed, for example, in Ref.~\cite{Arcadi:2021mag}. 

%For each of these models we determine the coupling that reproduces the observed relic density solving the full phase-space Boltzmann equation for the DM distribution function using the code DRAKE \cite{Binder:2021bmg}. Our aim is to reassess this well-studied class of models with a more complete treatment of the freeze-out dynamic for all the different models. 
For each model, we determine the portal coupling that reproduces the observed relic abundance and confront the resulting parameter space with the current LZ direct-detection limits~\cite{LZ:2024zvo}, the projected sensitivity of next-generation xenon experiments~\cite{DARWIN:2016hyl}, Fermi-LAT gamma-ray limits from dwarf spheroidal galaxies~\cite{Fermi-LAT:2016uux}, and the ATLAS bound on invisible Higgs decays~\cite{ATLAS:2023tkt}. 
We find that departures from kinetic equilibrium increase the predicted abundance below the resonance, by up to an order of magnitude at fixed coupling, thereby strengthening the direct-detection constraints. 
The effect is largest for the fermion with pseudoscalar interactions.
For each Higgs-portal model, we also investigate the parameter region where the setup constitutes only a fraction of the total DM abundance, rescaling the direct- and indirect-detection limits
consistently with predicted the abundance. 
%We also consider scenarios in which the Higgs-portal candidate constitutes only a fraction of the total DM abundance, rescaling the direct- and indirect-detection limitsconsistently with the abundance predicted by each treatment. 
% For each scenario, we compare the parameter region selected by the relic-density requirement with the most relevant experimental constraints. Direct-detection searches ~\cite{Schumann:2019eaa} provide the strongest bounds in large parts of the parameter space, since the Higgs portal generates spin-independent scattering between DM and nuclei. We therefore include the current LZ limits together with the projected sensitivity of DARWIN ~\cite{LZ:2024zvo,DARWIN:2016hyl}. Complementary information comes from indirect searches~\cite{Gaskins:2016cha}, in particular gamma-ray observations of dwarf spheroidal galaxies by Fermi-LAT~\cite{}, which constrain DM annihilation at late times. Finally, for DM masses below half of the Higgs mass, collider searches for invisible Higgs decays impose an additional constraint on the portal coupling~\cite{Boveia:2018yeb}, in particular we use the ATLAS constraint~\cite{ATLAS:2023tkt}.
% Finally we also consider the case that the DM candidate considered in this work makes up only a fraction of the total DM abbundance.

The paper is organized as follows. 
In Sec.~\ref{sec:Framework}, we introduce the Higgs-portal models, starting from their high-energy Lagrangians, and discuss their low-energy phenomenology relevant to relic-density calculations and experimental constraints. 
In Sec.~\ref{sec:relic}
%, we clarify the main differences with respect to previous studies. In particular, 
we compare the standard number-density Boltzmann equation with the full phase-space treatment used in this work
%, emphasizing 
and discuss the role of kinetic equilibrium near the Higgs resonance. 
In Sec.~\ref{sec:results}, we present the results for scalar, vector, and fermionic Higgs-portal DM. 
%For each case, we identify the regions of parameter space that reproduce the observed DM relic abundance and those in which the candidate accounts for only a fraction of the total DM density. 
%On top of this cosmological requirement, we impose the current experimental. 
We conclude in Sec.~\ref{sec:conclusion}.

%%%%%%%%%%%%%%%%%%%%%%%%%%%%%%%%%%%
\section{Higgs-Portal Dark Matter: Framework}
\label{sec:Framework}
We consider Higgs-portal DM models in which the DM candidate is either a real scalar $S$, an Abelian vector field $V_\mu$, or a Dirac fermion $\psi$. In general, these models can be described by
an effective field theory (EFT) Lagrangian of the form
\beq
\mathcal{L} = \mathcal{L}_{{\mathrm{SM}}}+\mathcal{L}_{\mathrm{P}}+\mathcal{L}_{\mathrm{D}},
\eeq 
where $\mathcal{L}_{\mathrm{SM}}$ denotes the Standard Model Lagrangian, $\mathcal{L}_{\mathrm{P}}$ denotes the portal interaction,
and $\mathcal{L}_{\mathrm{D}}$ describes interactions within the dark sector. We impose a discrete $\mathbb{Z}_2$ symmetry under which the DM fields are odd and all SM fields are even. This symmetry ensures the stability of the DM particle and forbids the otherwise possible kinetic mixing with hypercharge in the vector DM case.

Within this EFT framework, the portal interaction takes the form
\beq 
\mathcal{L}_{\mathrm{P}}\sim (\Phi^\dagger\Phi){\mathcal{O}}_{\mathrm{D}}, 
\eeq 
where $\Phi$ denotes the SM Higgs doublet and ${\mathcal{O}}_{\mathrm{D}}$ is a singlet operator of the lowest possible dimension constructed from the dark-sector field. For scalar and vector DM, ${\mathcal{O}}_{\mathrm{D}}$ is $S^2$ and $V_\mu V^\mu$, respectively, yielding dimension-four interactions with the SM Higgs field. For fermionic DM, both scalar, $\bar\psi\psi$, and pseudoscalar, $\bar\psi i\gamma_5\psi$, bilinears are possible, and the portal interaction has dimension five.  

%In this work we adopt a minimal and largely model-independent effective-field-theory (EFT)
%description of Higgs-portal dark matter (DM) \cite{Fedderke:2014wda}.
%The DM candidate is assumed to be either a real scalar $S$, a vector field $V_\mu$,
%or a Dirac fermion $\psi$.
Defining ${\mathcal{L}}_{\mathrm{DM}}\equiv {\mathcal{L}}_{\mathrm{P}}+{\mathcal{L}}_{\mathrm{D}}$, we write the Lagrangians for the scalar ($S$), vector ($V$), and fermionic ($\psi$) candidates explicitly as

\begin{align}
\mathcal{L}_S &=  \frac{1}{2} \partial_\mu S\partial^\mu S
%\partial_\mu S^\dagger \partial^\mu S 
- \frac{1}{2} \mu_S^2 S^2
%S^\dagger S  
- \frac{1}{4} \lambda_S S^4
%(S^\dagger S)^2 
- \frac{1}{2} \lambda_{HS} S^2\Phi^\dagger\Phi, 
% S^\dagger S \Phi^\dagger \Phi, 
\label{eq:LS} \\
\mathcal{L}_V &=  - \frac{1}{4} \mathcal{V}^{\mu\nu} \mathcal{V}_{\mu\nu} - \frac{1}{4} \lambda_V (V_\mu V^\mu)^2  + \frac{1}{2} \mu_V^2 V_\mu V^\mu + \frac{1}{2} \lambda_{HV}  V_\mu V^\mu \Phi^\dagger \Phi, \label{eq:LV} \\
\mathcal{L}_\psi &=  \bar{\psi} ( i \slashed{\partial} - \mu_\psi ) \psi - \frac{\lambda_{H\psi}}{\Lambda_\psi} \big( \cos\theta \, \bar{\psi} \psi + \sin\theta \, \bar{\psi} i \gamma_5 \psi \big) \Phi^\dagger \Phi. \label{eq:Lpsi}
\end{align}
In Eq.~\eqref{eq:LV}, $\mathcal{V}_{\mu\nu} = \partial_\mu V_\nu - \partial_\nu V_\mu$ is the field-strength tensor of vector DM. We adopt a Proca mass term for the vector field as a sufficient effective description for the present analysis. Because $V_\mu$ is not treated as a gauge field, the interactions in Eq.~\eqref{eq:LV} are subject to perturbative-unitarity constraints that become particularly restrictive at small DM masses~\cite{Lebedev:2011iq}. These constraints can be avoided in a UV completion in which vector DM is the gauge boson of a new $U(1)$ symmetry spontaneously broken by the vacuum expectation value of a dark Higgs field. The vector Higgs-portal interaction in Eq.~\eqref{eq:LV} is then recovered in the EFT limit of this setup~\cite{Arcadi:2020jqf,Arcadi:2023gox}.

\begin{comment}The scale $\Lambda_\psi$ appearing in Eq.~\eqref{eq:Lpsi} is associated with origin of the dimension-5 fermionic operator. This EFT interaction could arise from a more UV complete model where $\psi$ and $\Phi$ couple with another singlet scalar obtaining its vacuum expectation value and mass at scale $\Lambda_\psi\gg v$. 
\end{comment}
The scale $\Lambda_\psi$ appearing in Eq.~\eqref{eq:Lpsi} parametrizes the
origin of the dimension-five fermionic operator. In a UV-complete realization, this interaction may arise from integrating out additional degrees of freedom
coupled to $\psi$ and the Higgs sector. %In this work, however, we do not specify the UV completion and treat the operator phenomenologically, fixing $\Lambda_\psi=v$ in the numerical analysis.
Fermionic DM could alternatively be a Majorana particle, but we focus on the Dirac case because the phenomenology is generally similar.

After electroweak symmetry breaking (EWSB), the phenomenology of
%these models
the models in Eqs.~\eqref{eq:LS}--\eqref{eq:Lpsi} is fully
characterized by the DM mass and a small set of Higgs-portal couplings that control
DM annihilation, elastic scattering, and collider signatures. We now identify these parameters
for each DM candidate %described above 
separately.

%\begin{align}
%\mathcal{L}_S &= \mathcal{L}_{SM} + \frac{1}{2} \partial_\mu S^\dagger \partial^\mu S - \frac{1}{2} \mu_S^2 S^\dagger S - \frac{1}{2} \lambda_{HS} \Phi^\dagger \Phi S^\dagger S - \frac{1}{4} \lambda_S (S^\dagger S)^2, \label{eq:LS} \\
%\mathcal{L}_V &= \mathcal{L}_{SM} - \frac{1}{4} \mathcal{V}^{\mu\nu} \mathcal{V}_{\mu\nu} + \frac{1}{2} \mu_V^2 V_\mu V^\mu + \frac{1}{2} \lambda_{HV} \Phi^\dagger \Phi V_\mu V^\mu - \frac{1}{4} \lambda_V (V_\mu V^\mu)^2, \label{eq:LV} \\
%\mathcal{L}_\psi &= \mathcal{L}_{SM} + \bar{\psi} ( i \slashed{\partial} - \mu_\psi ) \psi - \frac{\lambda_{H\psi}}{\Lambda_\psi} \big( \cos\theta \, \bar{\psi} \psi + \sin\theta \, \bar{\psi} i \gamma_5 \psi \big) \Phi^\dagger \Phi. \label{eq:Lpsi}
%\end{align}

\subsection{Scalar and Vector Dark Matter}
After electroweak symmetry breaking, the Higgs doublet takes the unitary-gauge form
$\Phi \to (0,(v+H)/\sqrt{2})^{T}$, where $v \simeq 246~\mathrm{GeV}$ is the Higgs vacuum expectation value.
The Higgs-portal operators therefore generate both DM mass terms
and interactions with the physical Higgs boson.

For the real scalar field $S$, the Lagrangian in
Eq.~\eqref{eq:LS} becomes
\begin{equation}
\begin{aligned}
\mathcal{L}_S =
&\;\frac{1}{2} \partial_\mu S \partial^\mu S
- \frac{1}{2} m_S^2 S^2 - \frac{1}{4} \lambda_S S^4
- \frac{1}{2} \lambda_{HS} v H S^2
- \frac{1}{4} \lambda_{HS} H^2 S^2,
\end{aligned}
\label{eq:scalar_EWSB}
\end{equation}
where the physical scalar mass is
\begin{equation}
m_S^2 = \mu_S^2 + \frac{1}{2}\lambda_{HS} v^2 .
\label{eq:scalar_mass}
\end{equation}
The Higgs-portal interactions control DM annihilation into SM particles
and elastic scattering with the thermal bath, as well as the scattering processes
relevant for direct detection.
We assume the scalar self-coupling $\lambda_S$ is small enough so that the elastic collisions mediated by it can be neglected in the freeze-out calculations. We will therefore ignore $\lambda_S$ in what follows. 
% The scalar self-interaction term proportional to $\lambda_S$ is irrelevant to
% the observables considered here and will be neglected in what follows.

Similarly, after electroweak symmetry breaking, the vector-field Lagrangian in
Eq.~\eqref{eq:LV} takes the form
\begin{equation}
\begin{aligned}
\mathcal{L}_V =
&\; -\frac{1}{4} \mathcal{V}_{\mu\nu}\mathcal{V}^{\mu\nu}
+ \frac{1}{2} m_V^2 V_\mu V^\mu- \frac{1}{4} \lambda_V (V_\mu V^\mu)^2
+ \frac{1}{2} \lambda_{HV} v H V_\mu V^\mu
+ \frac{1}{4} \lambda_{HV} H^2 V_\mu V^\mu,
\end{aligned}
\label{eq:vector_EWSB}
\end{equation}
where the physical vector mass is
\begin{equation}
m_V^2 = \mu_V^2 + \frac{1}{2}\lambda_{HV} v^2 .
\label{eq:vector_mass}
\end{equation}
As in the scalar case, the Higgs interactions determine both the annihilation
rate relevant to the relic abundance and the elastic scattering rate with the thermal bath. Similar to the scalar case, we also assume $\lambda_V$ is small enough to be neglected in the analysis. 
%The vector self-interaction proportional to $\lambda_V$ does not play a significant role in this analysis and will be neglected.
%Finally, regarding the effective vector model, a renormalisable Lagrangian can be obtained by embedding it in a more complete framework, such as a simple UV completion with a dark Higgs sector that generates the vector mass through spontaneous symmetry breaking (SSB) \cite{Zaazoua:2021xls}. 

%%%%%%%%%%%%%%% FERMION %%%%%%%%%
%%%%%%%%%%%%%%%%%%%%%%%%%%%%%%%%%%%
\subsection{Fermionic Dark Matter}\label{sec:fermionicDM}
We now consider a Dirac fermion $\psi$ interacting with the SM through the dimension-five Higgs-portal operator introduced in Eq.~\eqref{eq:Lpsi}.
After EWSB, the Lagrangian becomes
\beq  
\mathcal{L}_\psi = \bar\psi i\slashed{\partial}\psi\hspace{-0.5pt}
-\hspace{-0.5pt}\Big(\mu_\psi \hspace{-0.5pt} +\hspace{-0.5pt} \frac{\tilde{\lambda}_{H\psi}v^2c_\theta}{2}\Big)\bar\psi\psi
- \frac{\tilde{\lambda}_{H\psi}v^2 s_\theta}{2}\bar\psi i\gamma_5\psi
- {\tilde{\lambda}_{H\psi}}\big(c_\theta\bar\psi\psi\hspace{-0.5pt}+\hspace{-0.5pt}s_\theta\bar\psi i\gamma_5\psi\big)\Big(vH\hspace{-0.5pt}+\hspace{-0.5pt}\frac{H^2}{2}\Big)~, \label{eq:fermion_after_EWSB}
\eeq
where $s_\theta\equiv\sin\theta$ and $c_\theta\equiv\cos\theta$. We have also defined the effective coupling $\tilde{\lambda}_{H\psi}\equiv \lambda_{H\psi}/\Lambda_\psi$, which has mass dimension $-1$.

Equation~\eqref{eq:fermion_after_EWSB} contains a pseudoscalar fermion mass term, so the mass is not yet in a purely scalar form. Following Ref.~\cite{Beniwal:2015sdl},
we move to the real-mass basis through a chiral rotation, $\psi \rightarrow \exp \left(i \gamma_5 \alpha / 2\right) \psi$ and $\bar{\psi} \rightarrow \bar{\psi} \exp \left(i \gamma_5 \alpha / 2\right)$, requiring the coefficient of the pseudoscalar mass bilinear $\bar{\psi} i \gamma_5 \psi$ to vanish. This fixes $\alpha$ through
\beq 
\tan \alpha=-\frac{\tilde{\lambda}_{H \psi}v^2 \sin \theta}{2}\left(\rule{0pt}{3ex}\right.\!\mu_{\psi}+\frac{\tilde{\lambda}_{H \psi}v^2 \cos \theta}{2}\left)^{}\rule{0pt}{3ex}\right.^{\!\!-1} .
\label{eq:chiralalpha}
\eeq
In the real-mass basis, the Lagrangian reads
\begin{align}
\label{eq:LpsiIR}
& \mathcal{L}_\psi=\bar{\psi} i \slashed{\partial} \psi-m_\psi \bar{\psi} \psi-\tilde{\lambda}_{H \psi}\left(\cos \xi \bar{\psi} \psi+\sin \xi \bar{\psi} i \gamma_5 \psi\right)\Big(v H+\frac{H^2}{2} \Big)~,
\end{align}
where $\xi \equiv \theta+\alpha$ and
\begin{align}
\cos \xi & =\frac{\mu_{\psi}}{m_{\psi}}\left(\rule{0pt}{3ex}\right.\!\!\cos \theta+ 
\frac{\tilde{\lambda}_{H \psi}v^2}{2 \mu_{\psi}}\!\left)\rule{0pt}{3ex}\right., \quad \sin \xi=\frac{\mu_{\psi}}{m_{\psi}} \sin \theta \label{eq:chiralxi}~, \\
\label{eq:chiralmpsi}
m_{\psi} & =\sqrt{\left(\rule{0pt}{3ex}\right.\!\!\mu_{\psi}+\frac{\tilde{\lambda}_{H \psi}v^2 \cos \theta}{2} \left)^{}\rule{0pt}{3ex}\right.^{\!\!2}+\left(\rule{0pt}{3ex}\right.\frac{\!\tilde{\lambda}_{H \psi}v^2 \sin\theta}{2}\left)^{}\rule{0pt}{3ex}\right.^{\!\!2}}~.
\end{align}

We consider three benchmark cases of fermionic DM, corresponding to different choices of the angle $\theta$ in the pre-EWSB Lagrangian~\eqref{eq:Lpsi}:

{\bf Case 1: $\theta =0$.}
The pre-EWSB Lagrangian~\eqref{eq:Lpsi} has a purely scalar coupling, and no pseudoscalar terms arise after EWSB. The post-EWSB Lagrangian is given by Eq.~\eqref{eq:LpsiIR}, with
\beq
\label{eq:case1_gps}
\cos\xi = 1~,\qquad \sin\xi = 0~.
\eeq

{\bf Case 2: $2\mu_\psi \cos\theta =-\tilde{\lambda}_{H\psi} v^2$.} The pre-EWSB Lagrangian~\eqref{eq:Lpsi} contains both scalar and pseudoscalar couplings. Equations~\eqref{eq:chiralxi} and \eqref{eq:chiralmpsi} yield $\xi = \pi/2$ (we discard the $\xi = -\pi/2$ solution, which would correspond to a negative effective coupling), and the post-EWSB Lagrangian~\eqref{eq:LpsiIR} has
\beq
\label{eq:case2_gps}
\cos\xi = 0~,\qquad \sin\xi = 1~.
\eeq

{\bf Case 3: $\theta =\pi/2$.}
The pre-EWSB Lagrangian~\eqref{eq:Lpsi} has a purely pseudoscalar coupling, but the post-EWSB Lagrangian~\eqref{eq:LpsiIR} contains both pseudoscalar and scalar couplings. Equations~\eqref{eq:chiralxi} and \eqref{eq:chiralmpsi} give
\beq
\label{eq:case3_gps}
\cos\xi = \frac{\tilde{\lambda}_{H\psi} v^2}{2 m_{\psi}}~,\qquad \sin\xi = \sqrt{1-\left(\rule{0pt}{3ex}\right.\frac{\tilde{\lambda}_{H\psi} v^2}{2 m_{\psi}}\left)^{}\rule{0pt}{3ex}\right.^{\!\!\!2}}~ ,
\eeq 
where we have again discarded solutions corresponding to negative effective couplings.

\section{Computing the relic density} 
\label{sec:relic}

The measured DM abundance  $\Omega_{\mathrm{DM}} h^2 = 0.120$ with percent-level uncertainty~\cite{Planck:2018vyg}.
%The DM abundance has been measured by the Planck Collaboration to be $\Omega_{\mathrm{DM}} h^2 = 0.120$, with percent-level uncertainty~\cite{Planck:2018vyg}.
In general, computing the relic density requires
solving the Boltzmann equation for the DM phase-space distribution $f_{\mathrm{DM}}(t,\mathbf{p})$: 
%in an expanding Friedmann--Robertson--Walker Universe~\cite{Kolb:1990vq}:
\begin{equation}
E \left( \partial_t - H \, \mathbf{p} \cdot \nabla_{\mathbf{p}} \right) f_{\mathrm{DM}}
= C[f_{\mathrm{DM}}]~.
\label{eq:fullBE}
\end{equation}
Here $E$ and $\mathbf{p}$ are the DM energy and momentum, respectively, and $H$ is the Hubble expansion rate.
The collision term has two contributions,
$C = C_{\mathrm{el}} + C_{\mathrm{ann}}$.
The elastic term $C_{\mathrm{el}}$ describes scattering off the thermal bath, ${\rm DM}\,f \to {\rm DM}\,f$, which changes the momenta of DM particles but not their number.
The annihilation term $C_{\mathrm{ann}}$ describes the inelastic processes ${\rm DM}\,{\rm DM} \leftrightarrow
{\rm SM}\,{\rm SM}$, which change the number of DM particles. Because the rate depends on the momenta of the annihilating particles, annihilation also changes the shape of $f_{\rm DM}$, rather than only its normalization.
%The collision operator $C$ encodes all interactions between DM and SM particles and can be decomposed into elastic and inelastic contributions,
%$C = C_{\mathrm{el}} + C_{\mathrm{ann}}$.
%Elastic scattering processes maintain kinetic equilibrium, while annihilation processes control chemical equilibrium.

The standard treatment assumes that elastic scattering maintains kinetic equilibrium between DM and the thermal bath throughout chemical decoupling.
This is usually well justified: the elastic scattering rate is proportional to the density of relativistic bath particles, whereas the annihilation rate is proportional to the Boltzmann-suppressed DM density. Therefore the elastic scattering rate can exceed the annihilation rate by many orders of magnitude at freeze-out, assuming the elastic scattering and the annihilation cross-sections remain of the same order of magnitude. 
The distribution therefore retains its equilibrium shape, $f_{\rm DM}(t,p) \propto f_{\rm DM,eq}(t,p)$, and
integrating Eq.~\eqref{eq:fullBE} over momentum yields the Zeldovich--Okun--Pikelner--Lee--Weinberg (ZOPLW) equation~\cite{Lee:1977ua} for the DM number density $n_{\mathrm{DM}}$.
Explicitly,
\begin{equation}
\frac{d n_{\mathrm{DM}}}{dt} + 3 H n_{\mathrm{DM}}
= - \langle \sigma v \rangle_T
\left( n_{\mathrm{DM}}^2 - n_{\mathrm{DM},\mathrm{eq}}^2 \right),
\label{eq:numberBE}
\end{equation}
where $n_{\mathrm{DM}} = g_{\mathrm{DM}} \int \! d^3p/(2\pi)^3 \, f_{\mathrm{DM}}(p)$ and $\langle \sigma v \rangle_T$ is the thermally averaged annihilation cross section~\cite{Gondolo:1990dk}.

% In the standard freeze-out scenario for weakly interacting massive particles, several simplifying assumptions are commonly adopted.
% Most importantly, it is often assumed that DM remains in kinetic equilibrium with the thermal bath throughout the chemical decoupling.
% Under this assumption, the phase-space distribution retains a thermal shape,
% $f_{\mathrm{DM}} \propto f_{{\mathrm{DM}},\mathrm{eq}}$,
% and Eq.~\eqref{eq:fullBE} reduces to an ordinary differential equation known as the Zeldovich--Okun--Pikelner--Lee--Weinberg Equation~\cite{Lee:1977ua}. For the number density $n_S$ we have:
% \begin{equation}
% \frac{d n_{\mathrm{DM}}}{dt} + 3 H n_{\mathrm{DM}}
% = - \langle \sigma v \rangle_T
% \left( n_{\mathrm{DM}}^2 - n_{\mathrm{DM},\mathrm{eq}}^2 \right),
% \label{eq:numberBE}
% \end{equation}
% Here $n_{\mathrm{DM}} = g_{\mathrm{DM}} \int \! d^3p/(2\pi)^3 \, f_{\mathrm{DM}}(p)$ and $\langle \sigma v \rangle_T$ denotes the thermally averaged annihilation cross section~\cite{Gondolo:1990dk}.

Near the Higgs resonance, $m_{\rm DM}\simeq m_h/2$, the assumption of kinetic equilibrium may fail~\cite{Binder:2017rgn}.
Annihilation through $s$-channel Higgs exchange is resonantly enhanced, allowing the observed abundance to be obtained with portal couplings much smaller than those required away from the resonance.
Elastic scattering, which proceeds through $t$-channel Higgs exchange, is not resonantly enhanced. It scales proportional to the small portal coupling and the Yukawa couplings of the bath particles, and  is  dominated by the heaviest fermions that remain abundant in the plasma.
% Instead, it is suppressed by the same small portal coupling and by the Yukawa couplings of the bath particles. It is therefore dominated by the heaviest SM fermions that remain abundant in the plasma.
Consequently, kinetic decoupling can occur before chemical decoupling as the annihilation term acts selectively in momentum space.
For $m_{\rm DM} < m_h/2$, only particle pairs with sufficient kinetic energy can reach the pole, $s \simeq m_h^2$, and 
annihilations deplete the
high-momentum tail of $f_{\rm DM}$. Once elastic scattering can no longer replenish this tail, the average DM energy falls below the bath temperature.
This reduces the fraction of pairs that can annihilate resonantly, making annihilation less efficient than assumed in Eq.~\eqref{eq:numberBE}.
The relic abundance is consequently larger than the standard prediction.
For $m_{\rm DM} > m_h/2$, the pole is approached from the
low-energy side of the distribution, and the deviations are smaller.

The degree to which kinetic equilibrium is violated depends on the Lorentz structure of the DM--Higgs coupling, which affects both the annihilation and elastic scattering rates.
For scalar and vector DM, annihilation is $s$-wave, and the
elastic amplitude is not suppressed by velocity at small momentum transfer. In these cases, the suppression of the elastic scattering comes entirely from the small coupling near the resonance. 
 
For fermionic DM, the impact of kinetic decoupling depends more strongly on the CP structure of the low-energy coupling in Eq.~\eqref{eq:LpsiIR}.
For a purely scalar coupling (case~1), Higgs-mediated annihilation
is $p$-wave suppressed.
Reproducing the observed abundance then requires larger portal couplings, which also enhance elastic scattering. DM therefore remains close to kinetic equilibrium during freeze-out even close to the resonance.
For a purely pseudoscalar coupling (case~2), annihilation is $s$-wave, and the observed abundance can be obtained with small couplings, as in the scalar and vector models.
Moreover, the elastic scattering amplitude is suppressed by the ratio of the momentum transfer to the DM mass, $|\mathbf{q}|/m_\psi$.
Kinetic decoupling therefore occurs earliest in this case, for which the deviation from the standard treatment is expected to be largest.
The mixed case~3 lies between these two limits. Annihilation is
dominated by the pseudoscalar coupling, whereas the unsuppressed elastic scattering contribution arises from the subdominant scalar coupling, $\cos\xi = \tilde\lambda_{H\psi}v^2/(2m_\psi)$.
We will see these features reflected in the results of Sec.~\ref{sec:results}.

% The assumption of kinetic equilibrium is usually well motivated, since elastic scattering
% rates with the thermal bath typically exceed annihilation rates by several orders of magnitude.
% \textcolor{blue}{The picture can be significantly altered if DM annihilation occur in the  resonance regime, $m_{\rm DM}\simeq m_h/2$. In this regime, the annihilation processes occurring via s-channel exchange of the mediator are significantly enhanced and the observed DM relic density is achieved for much smaller values of the portal coupling than for DM masses away from the resonance. Elastic scatterings of DM with the thermal plasma are not resonantly enhanced as they occur via t-channel exchange of the mediator. Therefore, the DM may fall out of kinetic equilibrium before the time of chemical freeze-out.}

To account for these effects, we compute the relic density with the \texttt{DRAKE} code~\cite{Binder:2021bmg}, which solves both the number-density equation and the full phase-space Boltzmann equation. Following the \texttt{DRAKE} notation, we denote the full Boltzmann-equation solution, Eq.~\eqref{eq:fullBE}, including both elastic and annihilation collision terms, by fBE. The approximate solution of Eq.~\eqref{eq:numberBE}, which assumes kinetic equilibrium, is denoted by nBE.

For the DM masses considered here, chemical freeze-out occurs at $T \sim m_{\rm DM}/25$, corresponding to a few GeV. The kinetic decoupling is sensitive to the details of the QCD transition as the elastic scattering rate depends on the quark content of the thermal bath. Because the Higgs couples to fermions proportionally to their masses, the rate is particularly sensitive to whether heavier quarks are treated as free scattering particles.
Following Ref.~\cite{Binder:2017rgn}, we bracket this uncertainty using the two limiting prescriptions implemented in \texttt{DRAKE}.
The first prescription, QCDA, maximizes the elastic scattering rate by treating all quark flavours as free down to $T_c=154\,{\rm MeV}$.
The second prescription, QCDB, provides a conservative estimate by retaining only the light-quark contributions, $u,d,s$, and only at temperatures above $4T_c\simeq600\,{\rm MeV}$.
The difference between these prescriptions provides an estimate of the theoretical uncertainty in the elastic collision term. Kinetic equilibrium persists longer with QCDA, so its results lie closer to those obtained with nBE.

% For the DM masses considered in this work, the temperature range relevant for decoupling is of order the GeV scale. In this regime, the elastic scattering
% rate can be sensitive to the treatment of the quark content of the thermal bath. Following Ref.~\cite{Binder:2017rgn}, we therefore estimate the associated uncertainty by considering the two limiting
% QCD prescriptions implemented in \texttt{DRAKE}.
% The first prescription, denoted QCDA, gives a maximal estimate of the elastic scattering rate by treating all quark flavours as deconfined down to $T_c=154\,{\rm MeV}$. The second prescription, QCDB, gives a conservative estimate by retaining only the light-quark contributions, $u,d,s$, and only
% for temperatures above $4T_c\simeq600\,{\rm MeV}$. The spread between these two treatments is interpreted as the theoretical uncertainty associated with the quark contribution to the elastic collision term.

\begin{figure}[H]
\centering
\includegraphics[width=0.7 \textwidth]{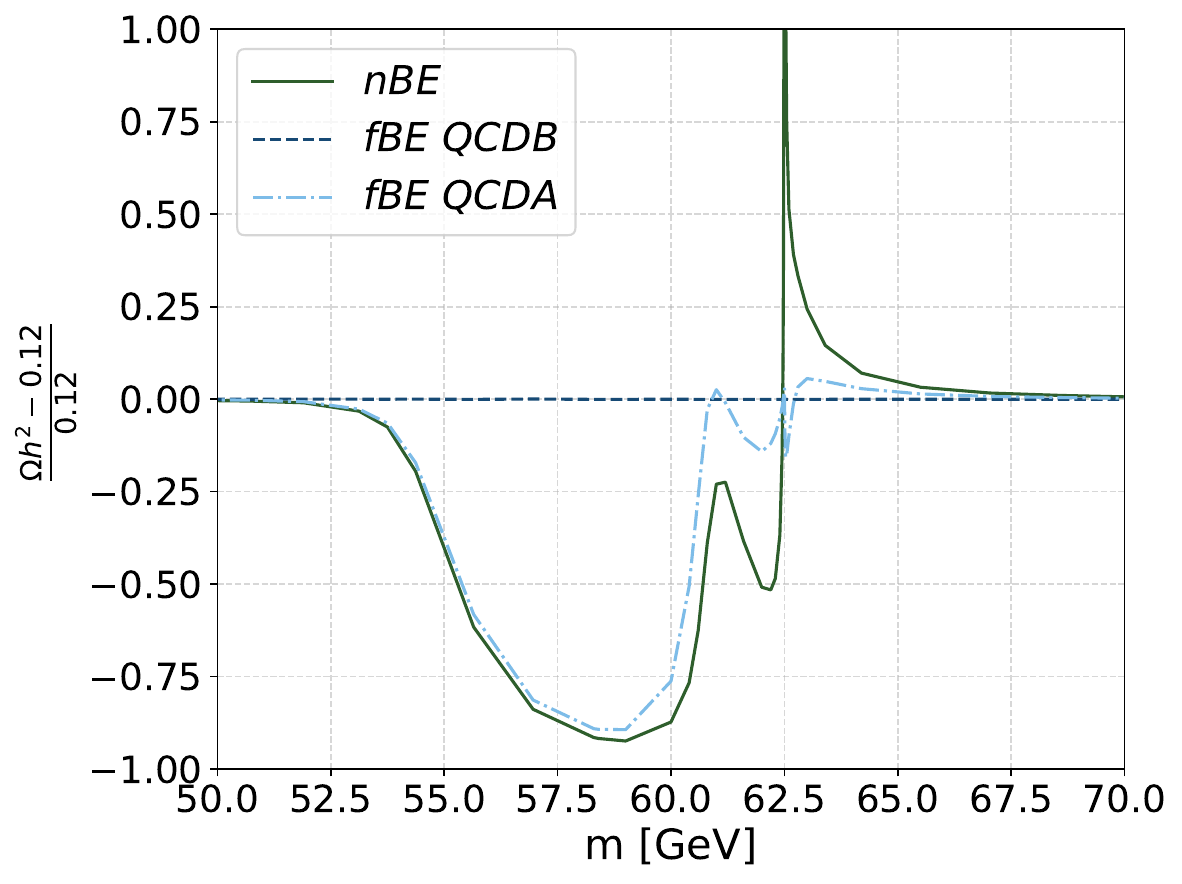}
\caption{Relative deviation $(\Omega h^2 - 0.12)/0.12$ of the predicted relic abundance from the observed value as a function of DM mass for fermionic DM with a purely pseudoscalar interaction after EWSB (case~2). For a given mass value, the coupling is fixed by requiring $\Omega h^2=0.12$ for the fBE QCDB solution (dark-blue dashed curve). For the same mass and coupling values, the abundance is then recomputed using the fBE QCDA (light-blue dot-dashed curve) and the nBE (solid green curve) solutions. 
%The dark-blue dashed curve shows fBE QCDB (zero by construction), the light-blue dot-dashed curve shows fBE QCDA, and the solid curve shows nBE.
%the observed relic density in the full Boltzmann treatment with the QCDB prescription. Keeping this mass and coupling fixed, the relic abundance is then recomputed using fBE QCDA and nBE. The dark-blue dashed, light-blue dot-dashed, and solid curves correspond to fBE QCDB, fBE QCDA, and nBE, respectively.
}
\label{fig:nBEfBE-diff}
\end{figure}

Fig.~\ref{fig:nBEfBE-diff} illustrates the difference between
the two treatments for fermionic DM with a purely pseudoscalar interaction after EWSB (case~2 of Sec.~\ref{sec:fermionicDM}).
For each DM mass, the portal coupling is chosen such that fBE with the QCDB prescription reproduces $\Omega h^2=0.12$. At fixed mass and coupling, we then recompute the abundance using fBE with QCDA and using nBE. Below the resonance, nBE underestimates the abundance by up to approximately 80\% relative to the QCDB result. The QCDA result lies between the other two curves and remains clearly separated from nBE, showing that the departure from kinetic equilibrium is robust against QCD uncertainties in this model. Away from the resonance, all three treatments agree.

% the solution of the full Boltzmann equation, Eq.~\eqref{eq:fullBE}, and Eq.~\eqref{eq:numberBE} for the DM relic density. The figure shows the curves for the fermion DM model where the parameters are fixed to correspond to purely pseudoscalar interaction after EWSB. The portal coupling at each value of DM mass is fixed so that the fBE solution with QCDB option, shown by the blue dashed line, yields the observed relic density. Using the same values values of the portal coupling, the fBE solution with QCDA option is shown with the light-blue dot-dashed line. The difference of these two quantifies the intrinsic systematic uncertainty of the computation. Finally, for the same values of the portal coupling, the solution of Eq.~\eqref{eq:numberBE} is shown as the solid line.  

%%%%%%%%%%%%%%%%%%%%%%%%%%%%%%%%%%%%%%%%%%%%%%%%%%%%%%%%%%%%%%%%%%%%%%%%%%%%%%%%%%%%%%%%%%%%%%%%%%%%%
%%%%%%%%%%%%%%%%%%%%%%%%%%%%%%%%%%%%%%%%%%%%%%%%%%%%%%%%%%%%%%%%%%%%%%%%%%%%%%%%%%%%%%%%%%%%%%%%%%%%%
%
\section{Results} 
\label{sec:results}
%
%%%%%%%%%%%%%%%%%%%%%%%%%%%%%%%%%%%%%%%%%%%%%%%%%%%%%%%%%%%%%%%%%%%%%%%%%%%%%%%%%%%%%%%%%%%%%%%%%%%%%
%%%%%%%%%%%%%%%%%%%%%%%%%%%%%%%%%%%%%%%%%%%%%%%%%%%%%%%%%%%%%%%%%%%%%%%%%%%%%%%%%%%%%%%%%%%%%%%%%%%%%

\begin{figure}
\centering

\begin{subfigure}[t]{0.48\textwidth}
    \centering
    \includegraphics[width=\textwidth]{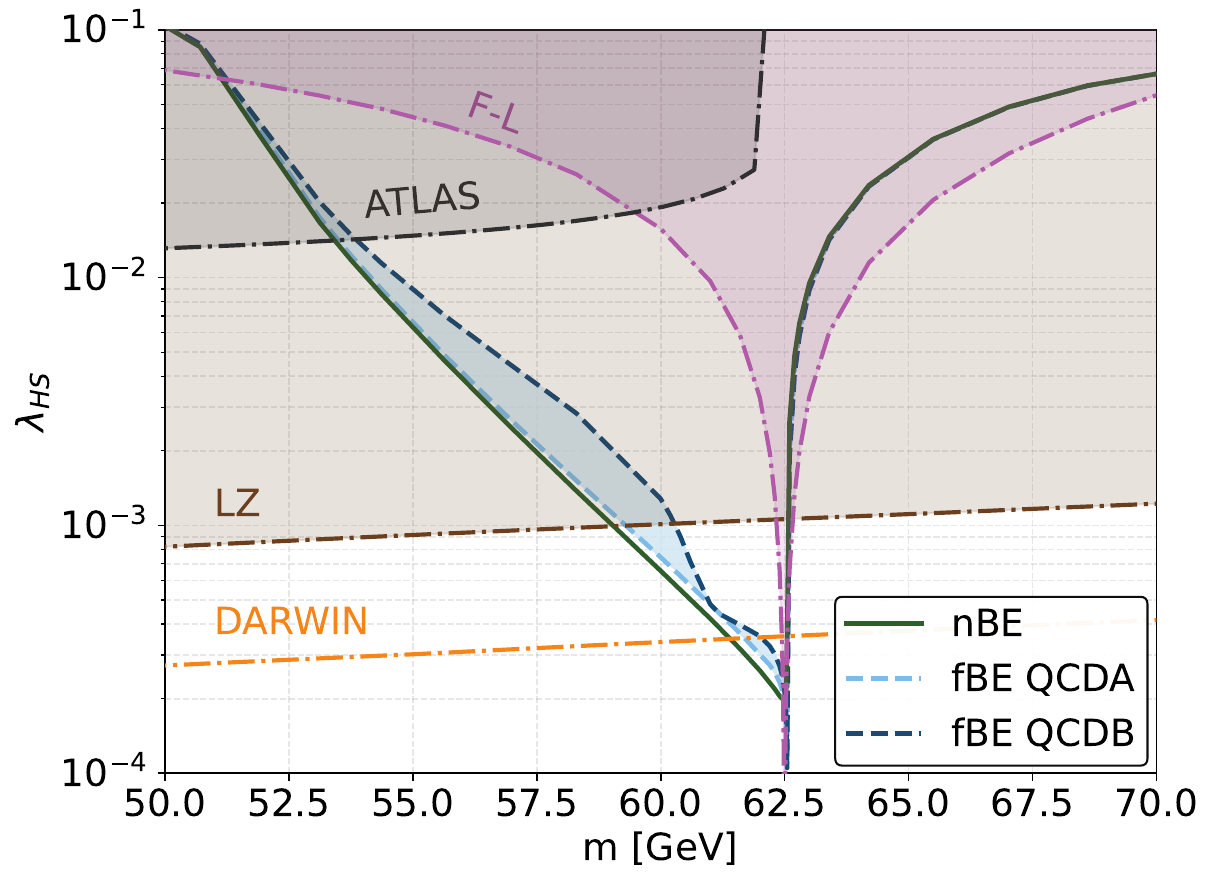}
\end{subfigure}
\hfill
\begin{subfigure}[t]{0.48\textwidth}
    \centering
    \includegraphics[width=\textwidth]{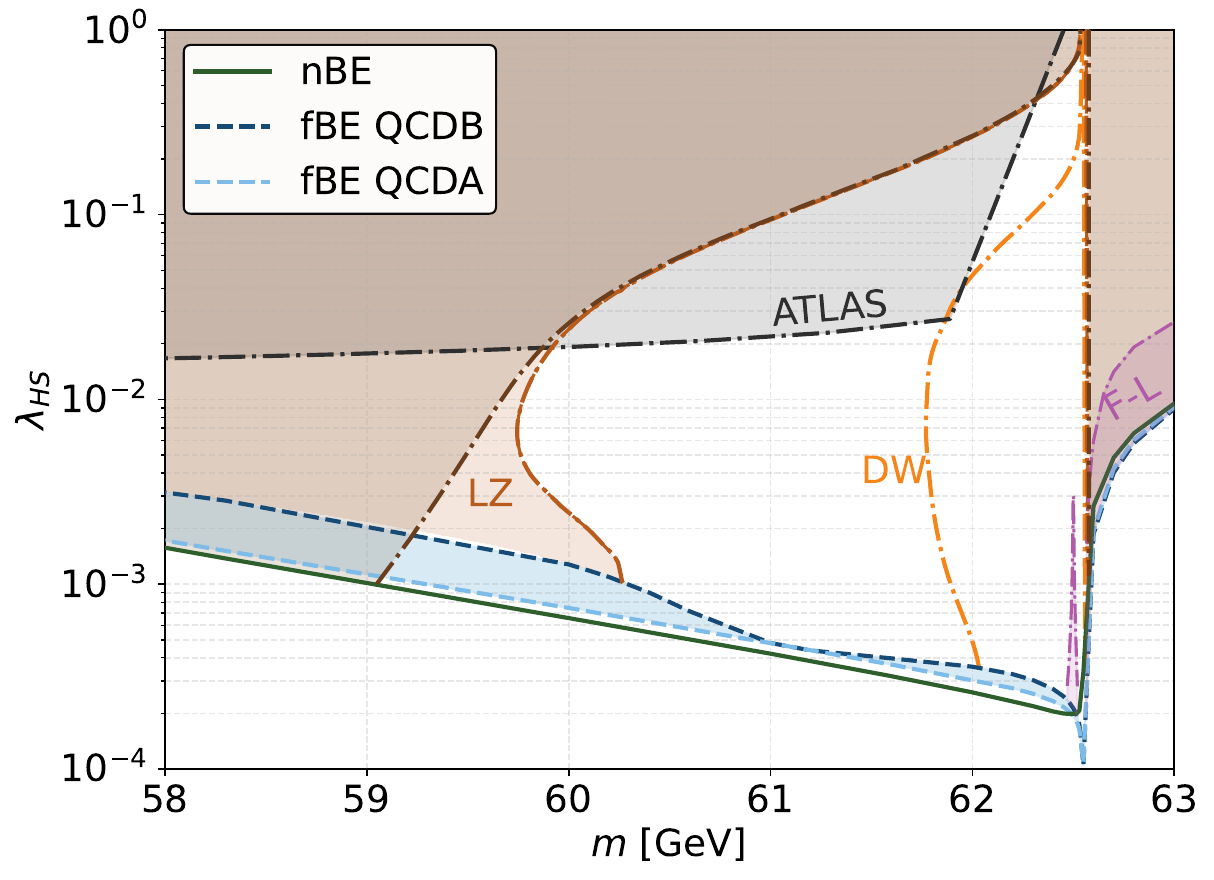}
\end{subfigure}

%\vspace{0.4cm}

\begin{subfigure}[t]{0.48\textwidth}
    \centering
    \includegraphics[width=\textwidth]{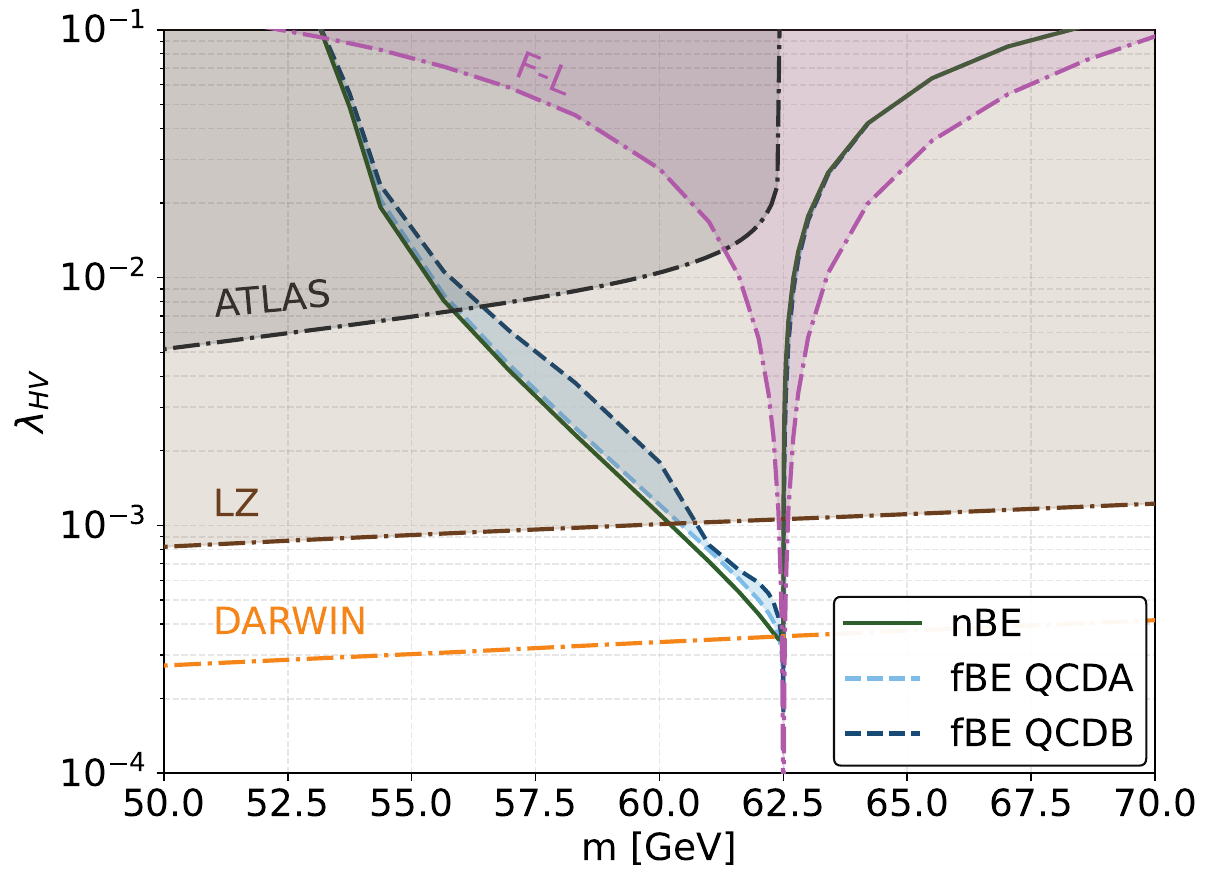}
\end{subfigure}
\hfill
\begin{subfigure}[t]{0.48\textwidth}
    \centering
    \includegraphics[width=\textwidth]{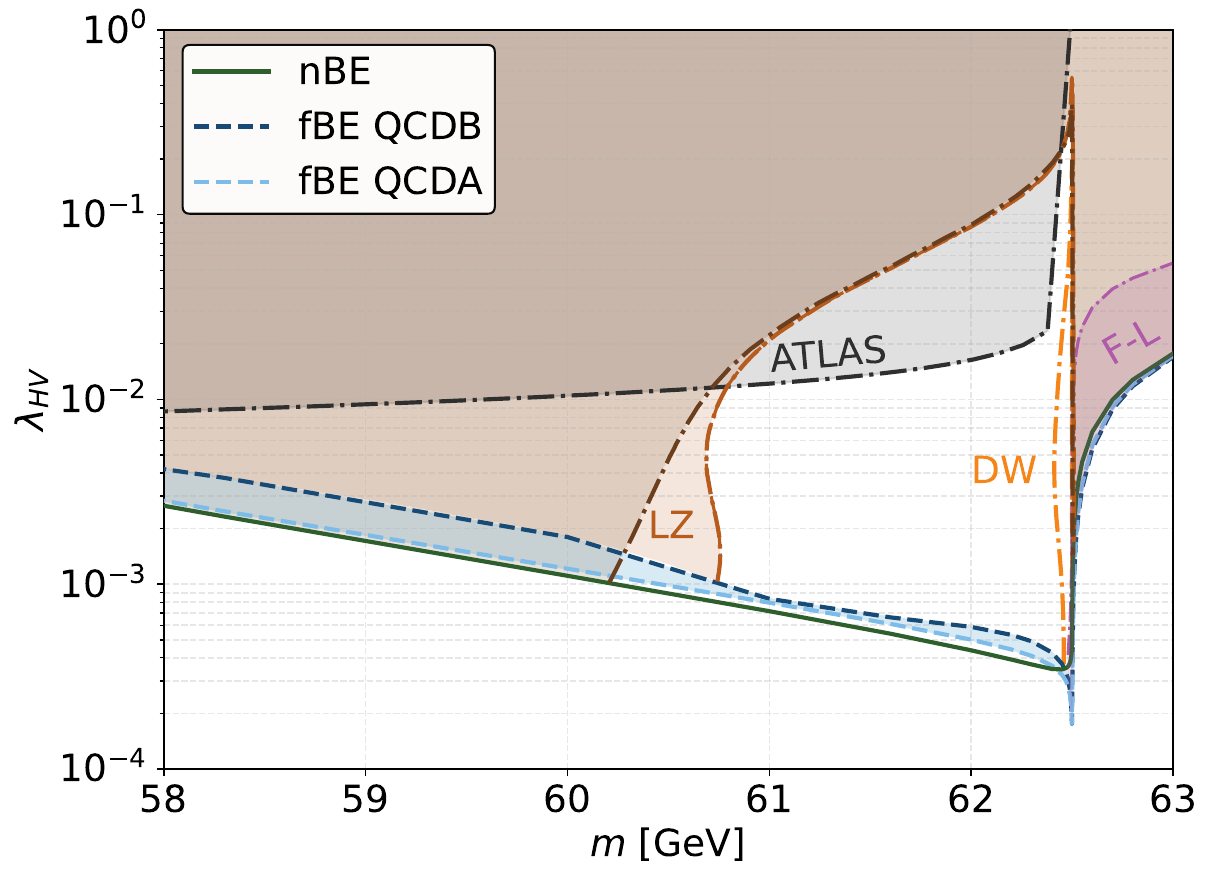}
\end{subfigure}

\caption{
Constraints on scalar and vector Higgs-portal DM in the resonant
region. The upper panels show the scalar model, and the lower panels show the
vector model. Solid green curves indicate the nBE relic-density solutions; dashed light-blue and dark-blue curves show the fBE results with the QCDA and QCDB prescriptions, respectively. Dash-dotted curves indicate the experimental constraints: LZ direct-detection bounds in brown, projected DARWIN (DW) sensitivity in orange, Fermi-LAT dwarf-spheroidal limits (F-L) in purple, and ATLAS invisible-Higgs-decay bounds in black. In the right panels, the direct-detection limits are rescaled by the fractional DM abundance, with $f_{\rm rel}$ obtained from the fBE QCDB solution for the grey dashed curve and from the nBE solution for the brown dashed curves.
}
\label{fig:scalar_vector}
\end{figure}

\begin{figure}
\centering
\includegraphics[width=0.7 \textwidth]{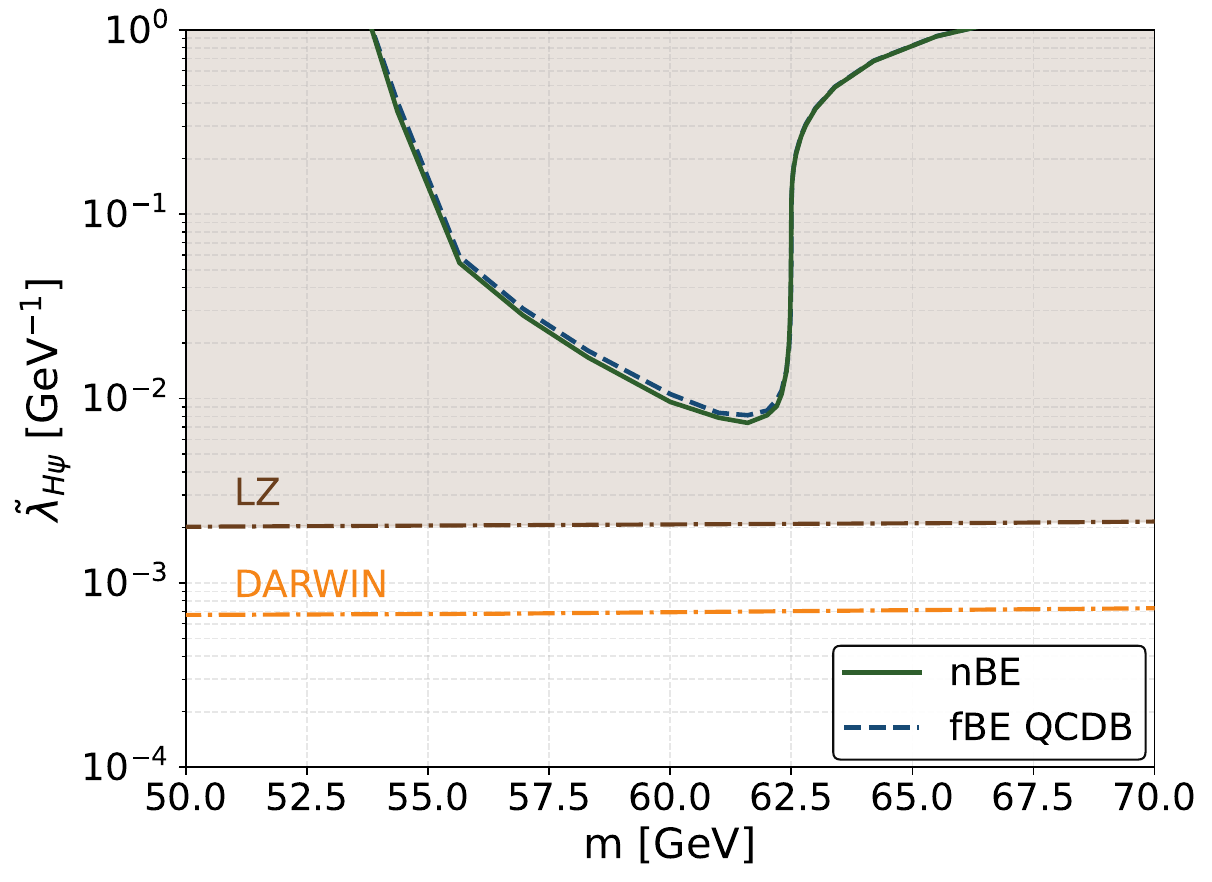}
\caption{Fermionic DM with a purely scalar interaction (case~1). Colours and line styles follow the conventions of the left panels of Fig.~\ref{fig:scalar_vector}.}
\label{fig:fermion_scalar}
\end{figure}

\begin{figure}
\centering

\begin{subfigure}[t]{0.48\textwidth}
    \centering
    \includegraphics[width=\textwidth]{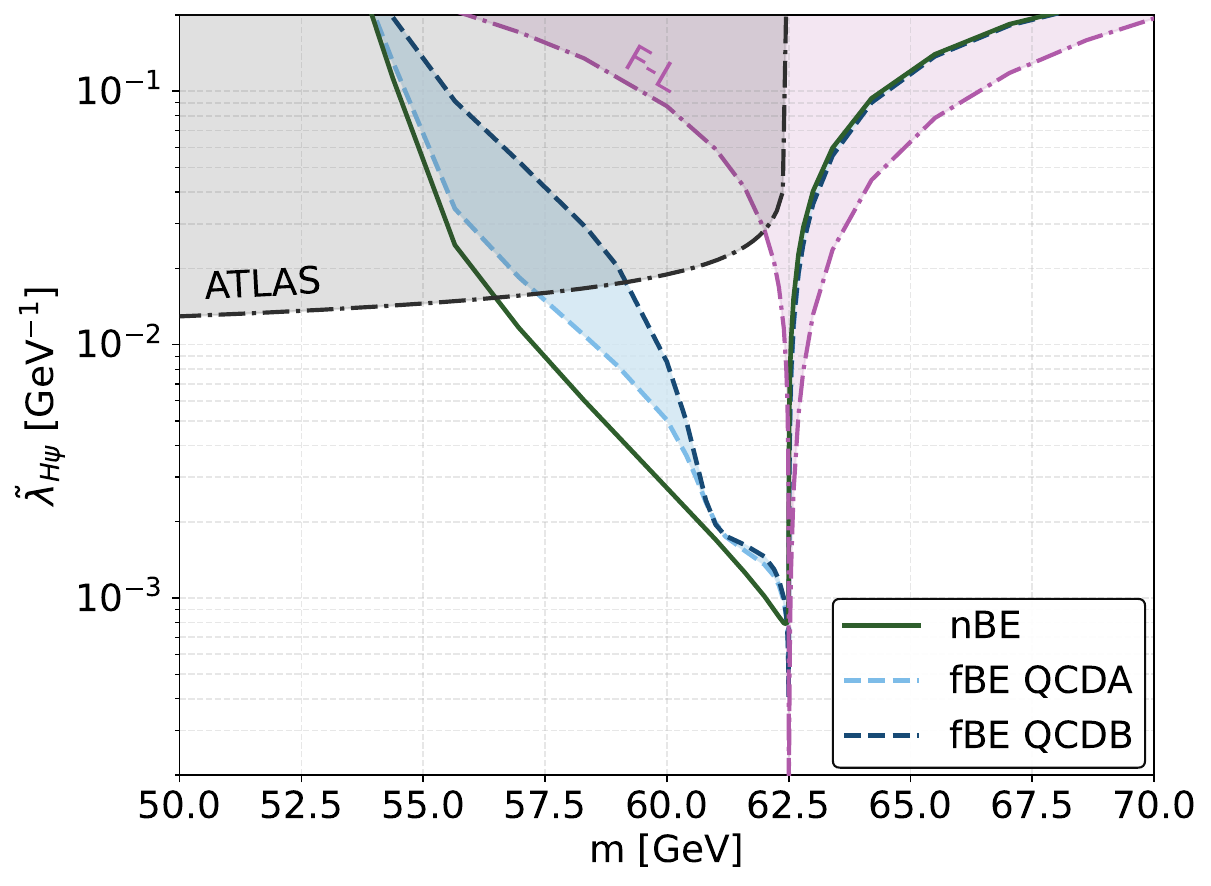}
\end{subfigure}
\hfill
\begin{subfigure}[t]{0.48\textwidth}
    \centering
    \includegraphics[width=\textwidth]{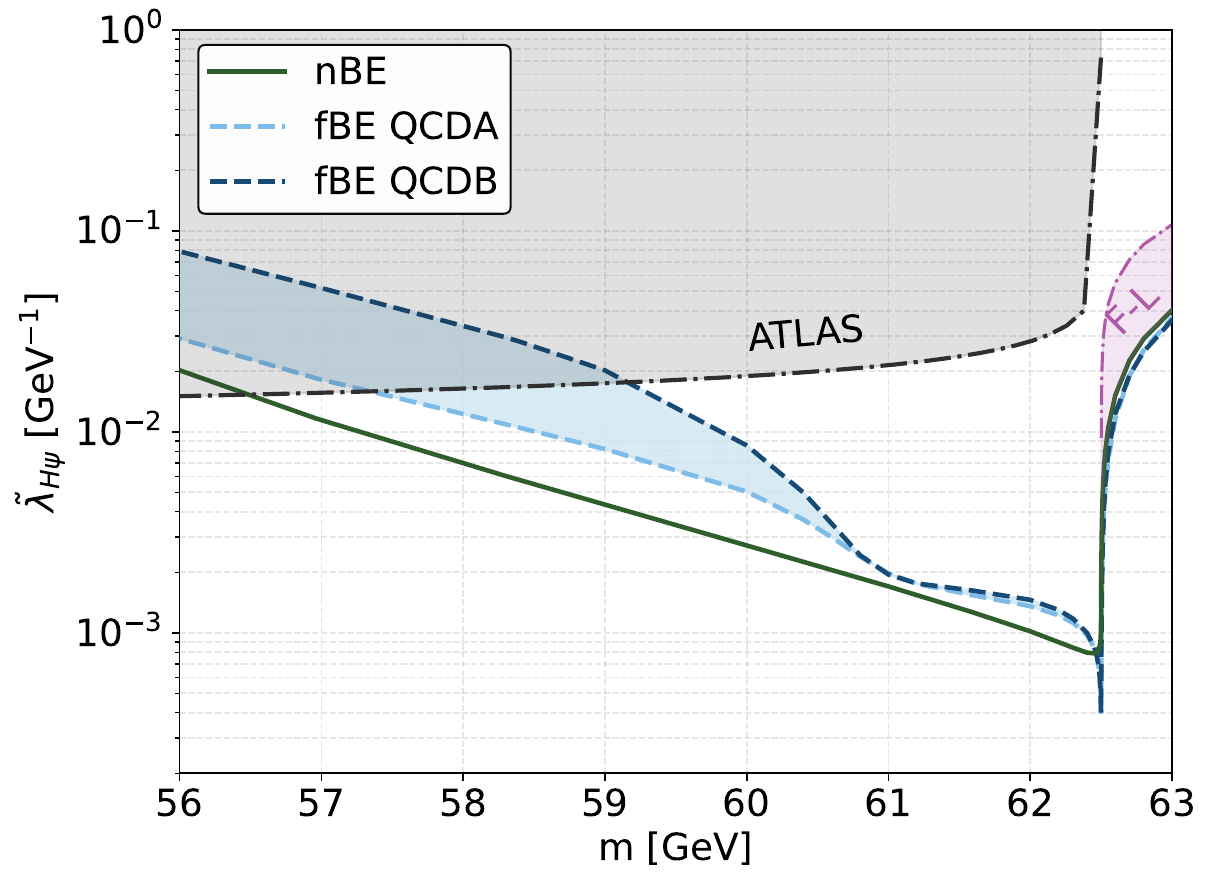}    
\end{subfigure}

\vspace{0.4cm}

\begin{subfigure}[t]{0.48\textwidth}
    \centering
    \includegraphics[width=\textwidth]{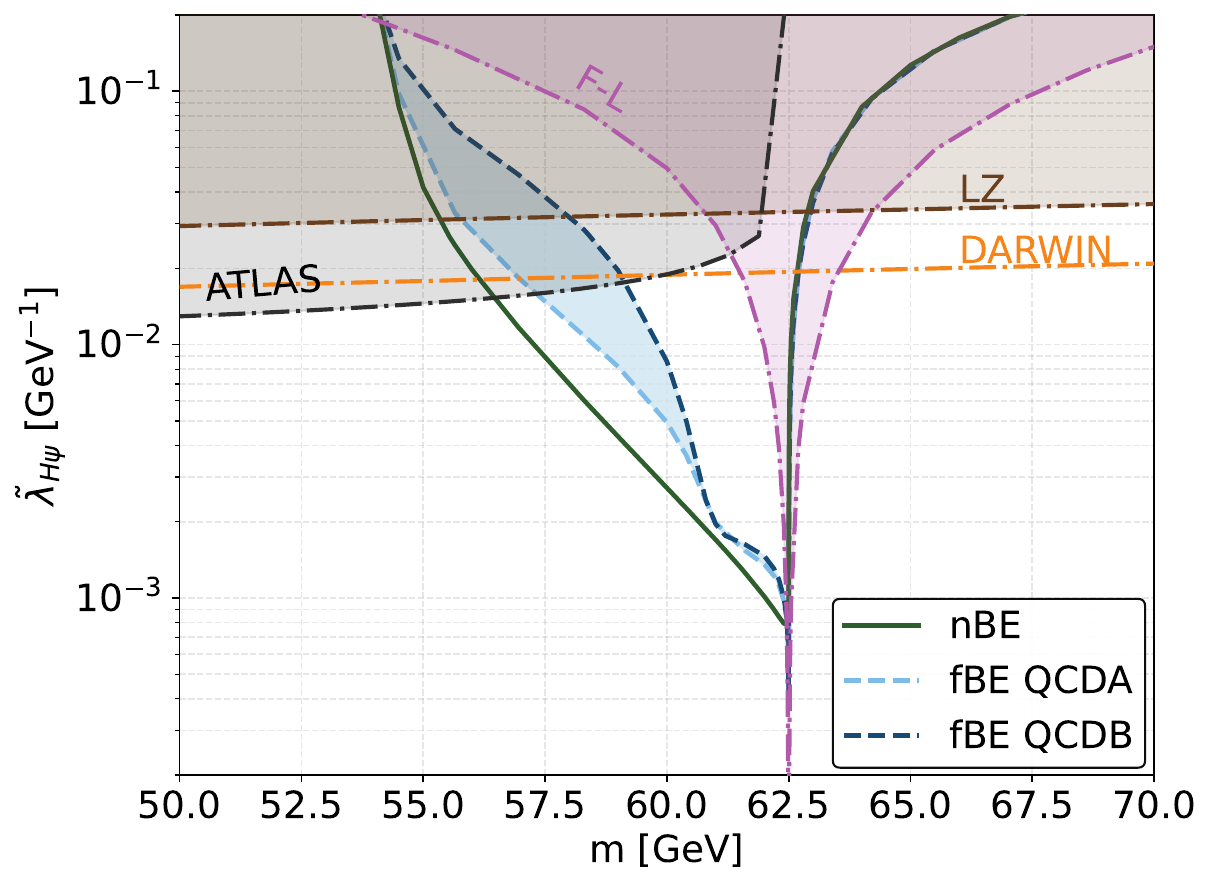}
\end{subfigure}
\hfill
\begin{subfigure}[t]{0.48\textwidth}
    \centering
    \includegraphics[width=\textwidth]{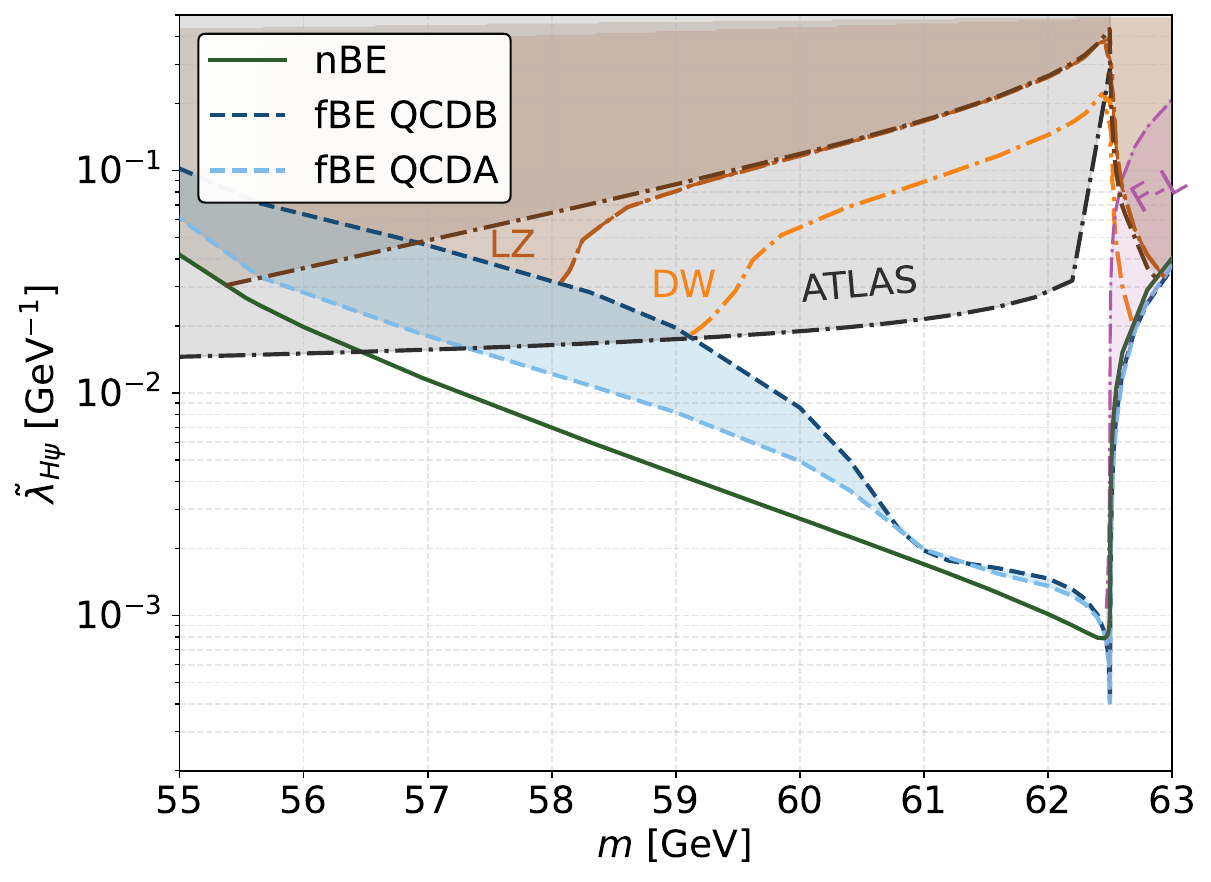}
    
\end{subfigure}

\caption{Constraints on fermionic Higgs-portal DM. The upper panels show the purely pseudoscalar interaction (case~2), and the lower panels show the mixed scalar--pseudoscalar interaction (case~3). The left panels assume $f_{\rm rel}=1$, while the right panels show the direct-detection constraints rescaled by the fractional DM abundance, using the fBE QCDB and nBE solutions. Colours and line styles follow the conventions of Fig.~\ref{fig:scalar_vector}.
}
\label{fig:fermion_case2_case3}
\end{figure}

In this section, we present our results for the Higgs-portal DM scenarios defined in Sec.~\ref{sec:Framework}. For each model, we have numerically solved the momentum-dependent Boltzmann equation to determine the relic abundance as a function of DM mass and portal coupling, correspondingly updated the direct- and indirect-detection limits, and imposed the collider constraints. Before discussing the results, we first describe how the curves in Figs.~\ref{fig:scalar_vector}, \ref{fig:fermion_scalar}, and \ref{fig:fermion_case2_case3} are obtained.

{\bf The relic abundance.} Curves labelled fBE and nBE indicate the DM abundances obtained by solving the full momentum-dependent Boltzmann equation and the integrated ZOPLW number-density equation, respectively. The latter assumes kinetic equilibrium and deviates from fBE near the resonance, $m_{\rm DM} \simeq m_h/2$. We parametrize the fraction of DM contributed by species $i=S,V,\psi$, corresponding to scalar, vector, or fermionic DM, as~\cite{Cline:2012hg}
	\begin{equation}
		f_{\rm rel} \equiv \frac{\Omega_{\rm DM,i} h^2}{0.12}~.  
		\label{eq:rescaling}
	\end{equation}
For $f_{\rm rel} = 1$, the candidate accounts for all of the observed DM; for $f_{\rm rel} < 1$, it constitutes only a fraction.
	
{\bf Direct-detection limits.} The direct-detection constraints shown in the figures are the current LZ bound~\cite{LZ:2024zvo} on the spin-independent DM--nucleon cross section, $\sigma_{\rm SI}$, and the projected DARWIN sensitivity~\cite{DARWIN:2016hyl}. We compute $\sigma_{\rm SI}$ with \texttt{micrOMEGAs}~\cite{Belanger:2013oya,Belanger:2006is}, using the required \texttt{CalcHEP}~\cite{Belyaev:2012qa} model files generated with \texttt{FeynRules}~\cite{Christensen:2008py,Alloul:2013bka}.
% We note that for $g_{\rm P}\neq0$ the cross-section contains a velocity suppressed extra term~\cite{Fedderke:2014wda} 
% % {\color{red}(SN: add some words to specify how this is obtained and why not implemented as the other cases)}{\color{blue}MB: For the fermonic case with $g_p\neq0$, the pseudo scalar interaction gives a velocity suppressed extra term which in the mixed case adds to the unsuppressed scalar contribution. Since the contribution vanishes in the zero momentum transfer limit, we do not evaluate it using the procedure adopted above. Instead following~\cite{Fedderke:2014wda}  , we integrate out the Higgs mediator and retain the leading term in the non relativisti expansion, obtaining:},
% 	\begin{equation}
% 		\sigma_{\rm SI}^{\psi N}
% 		\simeq
% 		\frac{C^2}{\pi}\,\mu_{\psi N}^2
% 		\left[
% 		\sin^2\alpha
% 		+\frac12\left(\frac{\mu_{\psi N}}{m}\right)^2 v^2
% 		\right],
% 		\qquad
% 		C=\frac{f_N \tilde{\lambda}_{H\psi}}{m_H^2}\, ,
% 		\label{eq:sigma_kolb_like}
% 	\end{equation}
% Here $\mu_{\psi N}=m_\psi \, m_N/(m_\psi+m_N)$ is the DM--nucleon reduced mass, $m_N$ is the nucleon mass, $m_h$ is the Higgs-boson mass, and $f_N \simeq 0.33\,\mathrm{GeV}$ denotes the effective Higgs--nucleon coupling~\cite{Agrawal:2010fh}. However, setting the  DM velocity $v = 220\,\rm{km/s}$, the velocity suppressed term vanishes in case 1 and is negligible in the other two cases for $\lambda_{H\psi}$ values compatible with direct detection constrains and $\Omega_{\rm DM}h^2\leqslant 0.12$. 
For fermionic DM, we note that the cross section is proportional to~\cite{Fedderke:2014wda}
	\begin{equation}
    \label{eq:sigma_SI_psiN}
		\sigma_{\rm SI}^{\psi N}
		\propto
		\cos^2\xi
		+\frac12\left(\frac{m_{N}}{m_\psi+m_N}\right)^2 v_{\rm DM}^2~,
	\end{equation}
% {\color{red}MB: in the reference and as I also did it was calculated for the pure pseudo-scalar case before EWSB. In order to have the complete formula (with $\xi$) probably we can just substitute $\sin{\alpha}\rightarrow\cos{\xi}$. The complete complete should be: \begin{equation}
% 		\sigma^{\rm SI}_{\psi N}
%         =
%         \frac{C^2}{\pi}\mu_{\psi N}^2\cos^2\xi
%         +
%         \frac{C^2}{2\pi}\mu_{\psi N}^2v^2\,\frac{M_N^2+M^2\cos^2\xi}{(M+M_N)^2}
% 	\end{equation}
% This is what I have found in the file $\rm{DM_nucleon_pseudoscalar.tex}$. Assuming that it is correct then in order to obtain the formula that kolb show we need to neglect
%  $v^2 \cos^2{xi}$ and then e have: \begin{equation}
% 		\sigma^{\rm SI}_{\psi N}
%         =
%         \frac{C^2}{\pi}\mu_{\psi N}^2\cos^2\xi
%         +
%         \frac{C^2}{2\pi}\mu_{\psi N}^2v^2\,\frac{M_N^2}{(M+M_N)^2} \propto cos^2\xi
%         +
%         \frac{1}{2}v^2\,\frac{M_N^2}{(M+M_N)^2}
% 	\end{equation}}\\
where $m_N$ is the nucleon mass and $v_{\rm DM}$ is the DM velocity in the nucleon rest frame. Taking $v_{\rm DM} = 220\,\rm{km/s}$, the velocity-suppressed term is negligible compared with the first term in cases~1 and~3, and we therefore neglect it in those cases. In case~2, it is the leading contribution, so the direct-detection constraints are strongly suppressed.

For $f_{\rm rel} < 1$, we assume that all DM components cluster in the same way in the Galaxy, so the local density of the Higgs-portal component is $\rho_{\rm DM,i}=f_{\rm rel} \rho_\odot$. Because the DM--nucleon scattering rate is proportional to the local density, the direct-detection constraint becomes $f_{\rm rel} \sigma_{\rm SI,i}\leq \sigma_{\rm SI}^{\rm tot}$, where $\sigma_{\rm SI}^{\rm tot}$ is the experimental upper limit obtained assuming that the candidate makes up all of the DM.
	
{\bf Indirect-detection limits.}
The gamma-ray flux from DM annihilation can be written as~\cite{Ambrogi:2018jqj}
	\beq
	\frac{d\Phi_\gamma}{dE_\gamma}
	=
	\frac{\langle\sigma v\rangle}{8 \pi \, m_{\rm DM}^2}
	\sum_j \, B_j \, \frac{dN_\gamma^j}{dE_\gamma} \, J \, ,
	\eeq	
where the sum runs over annihilation channels, $B_j=\langle\sigma v\rangle_j/\langle\sigma v\rangle$ is the branching fraction into final state $j$, and $dN_\gamma^j/dE_\gamma$ is the prompt photon spectrum per annihilation in that channel. The $J$ factor accounts for the DM density distribution along the line of sight.

We use the gamma-ray limits derived from Fermi-LAT (abbreviated as F-L in the figures) observations of dwarf spheroidal galaxies~\cite{Fermi-LAT:2016uux}. We evaluate the corresponding indirect-detection constraints with \texttt{MadDM}~\cite{Ambrogi:2018jqj}, using UFO model files~\cite{Darme:2023jdn, Degrande:2011ua} generated with \texttt{FeynRules}. For $f_{\rm rel} < 1$, the limits are rescaled by noting that
$J \propto \rho_{\rm DM, i}^2$, so the predicted flux scales as $f_{\rm rel}^2$.

{\bf Collider limits.} For $m_{\rm DM}<m_h/2$, the Higgs boson can decay invisibly into DM pairs, with partial widths given by
	\beq
	\begin{aligned}
		\Gamma_{H,\rm inv}(H \rightarrow SS)
		&=
		\frac{\lambda_{HS}^2 v^2}{32 \pi m_h}
		\left(1-\frac{4m_S^2}{m_h^2}\right)^{1/2},
		\\
		\Gamma_{H,\rm inv}(H \rightarrow VV)
		&=
		\frac{\lambda_{HV}^2 v^2 m_h^3}{128 \pi m_V^4}
		\left(
		1-\frac{4m_V^2}{m_h^2}
		+\frac{12m_V^4}{m_h^4}
		\right)
		\left(
		1-\frac{4m_V^2}{m_h^2}
		\right)^{1/2},
		\\
		\Gamma_{H,\rm inv}(H \rightarrow \bar{\psi}\psi)
		&=
		\frac{\tilde{\lambda}^2_{H\psi}m_h v^2}{8 \pi}
		\left(
		1-\frac{4 m_\psi^2}{m_h^2}
		\right)^{1/2}
		\left(
		1-\frac{4 m_\psi^2 {\rm cos}^{2}\xi}{m_h^2}
		\right).
	\end{aligned}
	\eeq
We impose the ATLAS 95\% confidence-level (C.L.) bound on the invisible Higgs branching ratio~\cite{ATLAS:2023tkt},
	\beq
	\mathrm{Br}_{\rm inv}
	=
	\frac{\Gamma_{H,\rm inv}}
	{\Gamma_{H,\rm inv}+\Gamma_{H,\rm SM}}
	<0.107~,
	\eeq 
where $\Gamma_{H,\rm SM}=4.1\,{\rm MeV}$~\cite{LHCHiggsCrossSectionWorkingGroup:2016ypw}.

%%%%%%%%%%%%%%%%%%%%%%%%%%%%%%%%%%%%%%%%%%%%%%%%%%%%%%%%%%%%%%%%%%%%%%%%%%%%%%%%%%%%
%FIG.2
%%%%%%%%%%%%%%%%%%%%%%%%%%%%%%%%%%%%%%%%%%%%%%%%%%%%%%%%%%%%%%%%%%%%%%%%%%%%%%%%%%%%
%%%%%%%%%%%%%%%%%%%%%%%%%%%%%%%%%%%%%%%%%%%%%%%%%%%%%%%%%%%%%%%%%%%%%%%%%%%%%%%%%%%%
%%%%%%%%%%%%%%%%%%%%%%%%%%%%%%%%%%%%%%%%%%%%%%%%%%%%%%%%%%%%%%%%%%%%%%%%%%%%%%%%%%%%

{\bf Results.} Fig.~\ref{fig:scalar_vector} shows the results for scalar DM (upper row) and vector DM (lower row). In the left panels, the direct- and indirect-detection constraints are calculated assuming $f_{\rm rel}=1$. In the right panels, the constraints are rescaled using $f_{\rm rel}(m,\lambda)$ evaluated at each point in the mass--coupling plane and are shown only in the region $f_{\rm rel}\leqslant 1$. The curves labelled fBE QCDA and fBE QCDB show the $f_{\rm rel}=1$ contours obtained from the full momentum-dependent Boltzmann equation, while the nBE curve shows the corresponding contour from the ZOPLW equation.

For the scalar-portal model, the momentum-dependent Boltzmann equation has been studied previously in \cite{Binder:2017rgn,Binder:2021bmg,DiMauro:2023tho}, and our results agree well with these studies. For the vector and fermionic portals, we are not aware of previous comprehensive investigations using the full momentum-dependent treatment. 

Fig.~\ref{fig:scalar_vector} shows that departures from kinetic equilibrium during freeze-out produce qualitatively similar differences between the full and nBE solutions for scalar and vector DM. In both cases, at fixed portal coupling, nBE underestimates the DM abundance for masses below the resonance. The quantitative difference depends on which quark species are included as relativistic degrees of freedom in the thermal bath, as illustrated by the shaded area between the fBE QCDA and QCDB curves.
% In terms of $f_{\rm rel}$ the difference ranges from {\color{blue} {\cal O}(XX)\%} for QCDA to {\color{blue} {\cal O}(XX)\%} for QCDB in the scalar case 
% {\color{red} MB: If you mean the relative difference $(\Omega_{\rm nBE}-0.12)/0.12$, evaluated at the portal coupling for which the fBE solution gives $\Omega h^2=0.12$ for QCDA and QCDB separately, we find a peak in the maximum deviations of approximately $920\%$ for QCDB and $877\%$ for QCDA, both at a DM mass of $62.55$ GeV (the same that we see in figure 1 that is cut but the peak also there reaches order 10)}. 
% %MICHELE: Let's keep extended discussions as comments rather than coloured text in the output file. Otherwise things get too cluttered.  
% In the vector case the differences are somewhat smaller, ranging from {\color{blue} {\cal O}(XX)\%} for QCDA to {\color{blue} {\cal O}(XX)\%} for QCDB. 
% %{\color{red}MB: Do you mean the differences plotted in fig 1 but for both the scalar and vector case?}. 

The full phase-space treatment also leads to stronger direct-detection constraints than nBE, which underestimates the DM abundance. In Fig.~\ref{fig:scalar_vector}, we show the LZ constraints obtained using the fBE QCDB and nBE solutions. The limits obtained using fBE QCDA would be close to the nBE limits for both scalar and vector DM and are omitted to avoid clutter. The shaded region between the two LZ curves therefore illustrates the uncertainty associated with the different QCD prescriptions.
We also show the projected sensitivity of the planned DARWIN experiment, computed using the fBE QCDB solution, as an orange dot-dashed curve. This sensitivity is approximately a factor of three stronger than the current LZ bounds.

The indirect-detection constraints are most relevant for masses above the resonance. Although the fBE and nBE solutions also differ in this region, as shown in Fig.~\ref{fig:nBEfBE-diff}, $f_{\rm rel}$ depends so steeply on $\lambda$ that these differences shift the indirect-detection curves only slightly. The Fermi-LAT (F-L) limits in the right panels of Fig.~\ref{fig:scalar_vector} are calculated with fBE QCDB; using fBE QCDA or nBE would produce no appreciable change. This also holds for the peak in the F-L curve for scalar DM at the resonant mass, just below the dip in the $f_{\rm rel}=1$ contour. The peak results from the interplay between resonant enhancement of the annihilation rate relevant to indirect detection and the steep dependence of $f_{\rm rel}$ near the resonance. In addition to the resonance enhancement, the latter is sensitive to the details of freeze-out, which shift the dip in the $f_{\rm rel}=1$ curve slightly above the resonance in the scalar case. For vector DM, the dip instead occurs slightly below the resonance, and the corresponding peak in the F-L curve is almost entirely washed out.

%%%%%%%%%%%%%%%%%%%%%%%%%%%%%%%%%%%%%%%%%%%%%%%%%%%%%%%%%%%%%%%%%%%%%%%%%%%%%%%%%%%%
%%%%%%%%%%%%%%%%%%%%%%%%%%%%%%%%%%%%%%%%%%%%%%%%%%%%%%%%%%%%%%%%%%%%%%%%%%%%%%%%%%%%
Fig.~\ref{fig:fermion_scalar} shows the results for fermionic case~1, corresponding to a purely scalar coupling. This case is completely excluded by the current direct-detection constraints. The difference between the fBE QCDB and nBE solutions is negligible (as is the difference between fBE QCDA and nBE, although QCDA is not shown), demonstrating that departures from kinetic equilibrium have little impact on freeze-out. The LZ bound shown in the figure assumes $f_{\rm rel}=1$. However, because the bound lies below the contour on which the model yields $f_{\rm rel}(m,\tilde{\lambda}_{H\psi})=1$, case~1 is also excluded as a subdominant DM component.

%%%%%%%%%%%%%%%%%%%%%%%%%%%%%%%%%%%%%%%%%%%%%%%%%%%%%%%%%%%%%%%%%%%%%%%%%%%%%%%%%%%%
%FIG.4
%%%%%%%%%%%%%%%%%%%%%%%%%%%%%%%%%%%%%%%%%%%%%%%%%%%%%%%%%%%%%%%%%%%%%%%%%%%%%%%%%%%%

%%%%%%%%%%%%%%%%%%%%%%%%%%%%%%%%%%%%%%%%%%%%%%%%%%%%%%%%%%%%%%%%%%%%%%%%%%%%%%%%%%%%
%%%%%%%%%%%%%%%%%%%%%%%%%%%%%%%%%%%%%%%%%%%%%%%%%%%%%%%%%%%%%%%%%%%%%%%%%%%%%%%%%%%%

Fig.~\ref{fig:fermion_case2_case3} shows the results for fermionic case~2 (upper row) and case~3 (lower row). The curves are calculated as described for Fig.~\ref{fig:scalar_vector}. In both cases, the differences between fBE and nBE are larger than those in the scalar and vector models. Thus, departures from kinetic equilibrium affect freeze-out more strongly in these fermionic scenarios than in the scalar and vector models, unlike fermionic case~1 discussed above. In particular, the fBE solution differs visibly from nBE even with the QCDA prescription.

Fermionic case~2 corresponds to a purely pseudoscalar coupling. Here the leading contribution to the DM--nucleon cross section in Eq.~\eqref{eq:sigma_SI_psiN} is velocity dependent and strongly suppressed in the nonrelativistic limit. As a result, the direct-detection bounds do not constrain the parameter range shown in Fig.~\ref{fig:fermion_case2_case3}. In case~3, the fermion has a dominant pseudoscalar coupling and a subdominant scalar coupling; see Eq.~\eqref{eq:case3_gps}. Even this subdominant scalar contribution makes the direct-detection constraints relevant. As in the scalar and vector models, departures from kinetic equilibrium strengthen these bounds.

\section{Conclusion}\label{sec:conclusion}
The resonance region $m_{\rm DM} \simeq m_h/2$ remains one of the few viable regions in the parameter space of the minimal realizations of the Higgs portal model. We have revisited the computation of the DM thermal relic density by solving the full phase-space Boltzmann equation via the publicly available package \texttt{DRAKE}. The parameter space accounting for the correct relic density has been subsequently compared with the current and projected DM direct indirect detection constraints. For the scalar, vector and fermion Higgs portal models we have considered in this work, we found that the standard treatment, based on the integrated Boltzmann equation, always underestimates the relic abundance below the resonance. The quantitative difference depends on the DM spin. 

The scalar and vector DM models show a qualitatively similar behaviour. 
For DM masses below the resonance, our results show that the couplings required to reproduce the observed relic abundance are about 2 times larger than indicated by the standard treatment.  

In the case of fermion DM, the CP structure of the coupling plays a crucial role. For a purely CP conserving portal coupling the difference between the full momentum dependent result and the standard treatment remains negligible for all mass values. 
For a purely CP violating coupling the conclusion is opposite: the difference is significant, up to an order of magnitude in terms of the coupling.

The larger couplings required for obtaining correct relic density lead to increased tension with constraints from direct detection for scalar and vector Higgs portal models. For fermion DM the CP-conserving coupling is ruled out by the LZ constraint. In the case with purely CP-violating coupling the direct detection constraints are negligible due to the suppression of the scattering rate with the momentum transfer. In the mixed case, even the subdominant scalar coupling makes direct detection relevant. 

Indirect-detection bounds are mainly important above the resonance and are relatively insensitive to the treatment of kinetic decoupling because the relic abundance depends steeply on the coupling. 
Finally, since our results push the required couplings towards 
larger values, the constraint from invisible decays of the Higgs becomes more relevant. Especially for the fermion dark matter this provides currently the most stringent constraint of the model below the resonance.

Our results show that the full momentum-depedent analysis is necessary to determine the predictions of Higgs portal DM models near the resonance. Compared to the standard treatment,  the full results shift the viable mass and coupling region and strengthen the direct detection constraints. The fermionic case with a pure scalar coupling is already ruled out by the current direct detection constraints. The sensitivity of future xenon-based experiments can exclude most of the remaining parameter space of the scalar and vector models. The fermionic cases with a pseudoscalar coupling studied in this work will remain testable mainly through the invisible Higgs decays and indirect searches. Future Higgs factories and gamma-ray observations will improve the sensitivity of these probes. We note that while we have not studied the case with Majorana fermions separately, the results for that case should be qualitatively similar to the Dirac setups. A natural extension of this work would be to embed the vector and fermion effective theories into explicit UV completions, in which additional scalar states could modify both the resonant annihilation and kinetic decoupling dynamics.

\section{Acknowledgments}
MDM acknowledges support from the research grant {\sl TAsP (Theoretical Astroparticle Physics)} funded by Istituto Nazionale di Fisica Nucleare (INFN). We also thank Andrzej Hryczuk for useful discussions concerning the use of the \texttt{DRAKE} package.

 \appendix
\section{Relevant matrix elements and cross sections}
\label{app:cross_sections}
Except for the channel with an $hh$ final state, DM annihilation into SM particles
proceeds through an $s$-channel Higgs mediator. The annihilation cross section
multiplied by the M{\o}ller velocity can be expressed as
\beq
\sigma_i v_{\rm Mol}
=
2\,\frac{
	\overline{|\mathcal{M}_{{\rm DM}{\rm DM}\rightarrow H}|}^2
}{\sqrt{s}}\,
|D_H(s)|^2\,
\Gamma_{H\rightarrow i}(\sqrt{s}) .
\label{eq:sigmav_common}
\eeq
Here $\overline{|\mathcal{M}_{{\rm DM}{\rm DM}\rightarrow H}|}^2$ is the squared matrix element for producing the intermediate Higgs boson, averaged over the initial-state degrees of freedom and summed over the final-state ones. The quantity $\Gamma_{H\rightarrow i}(\sqrt{s})$ is the partial width for the decay of a Higgs boson of invariant mass $\sqrt{s}$ into final state $i$. The Higgs propagator is defined as
\beq
|D_H(s)|^2 =
\frac{1}{(s-m_h^2)^2 + m_h^2 \left(\Gamma_H^{\rm tot}\right)^2} ,
\label{eq:higgs_propagator}
\eeq
where $\Gamma_H^{\rm tot}$ is the total Higgs width, including both SM decay channels and invisible decays into DM pairs when kinematically allowed:
\beq
\Gamma_H^{\rm tot}
=
\Gamma_{H,\rm SM}
+
\Gamma_{H,\rm inv} .
\eeq
In the numerical analysis, we use $\Gamma_{H,\rm{SM}}=4.1\,{\rm MeV}$~\cite{LHCHiggsCrossSectionWorkingGroup:2016ypw}.
%The $hh$ final state is not included, since it is kinematically closed in the mass range considered in this work.

%To solve the full phase-space Boltzmann equation, elastic scattering between DM and SM particles must also be included. In DRAKE, this is implemented through the quantity:
%\beq \int_{-4 k_{\rm cm}^2}^{0} {\rm d}t\,(-t)\, |\mathcal{M}_{{\rm DM}\, X \leftrightarrow {\rm DM}\, X}|^2 , \label{eq:tM2_def} 
%\eeq
%where $X=f$ denotes a SM fermion and $X=B$ denotes a massive vector boson ($B=W,Z$). Here $k_{\rm cm}$ is the momentum in the center-of-mass frame and the squared matrix element is summed over final spins or polarizations, but not averaged over the initial state (for more details, see \cite{Binder:2017rgn}).

For the different DM scenarios, the squared matrix elements entering Eq.~\eqref{eq:sigmav_common} are
\begin{flalign} 
	\overline{|\mathcal{M}_{SS\rightarrow H}|}^2 &= (\lambda_{HS} v)^2 ,\\
	\overline{|\mathcal{M}_{VV\rightarrow H}|}^2 &=\frac{ (\lambda_{HV} v)^2 }{36 \, m_V^4} (12 \, m_V^4 + s^2-4\, s\, m_V^2) ,\\
	\overline{|\mathcal{M}_{\psi \overline{\psi}\rightarrow H}|}^2 &=\frac{ (\tilde{\lambda}_{H\psi} v)^2 }{2 } (s-4 \, m_\psi^2 \, \cos{\xi}^2).
\end{flalign}
Recalling that $\xi=\alpha+\theta$, the fermionic DM expression reduces to the three benchmark scenarios introduced in Sec.~\ref{sec:fermionicDM} when the corresponding relations among $\theta$, $\alpha$, and $\xi$ are imposed.

Solving the full phase-space Boltzmann equation also requires elastic scattering between DM and SM particles. In \texttt{DRAKE}, the model-dependent input to the elastic collision term is given by the squared matrix elements for
$
{\rm DM}\, X\leftrightarrow{\rm DM} \, X
$,
where $X=f$ denotes an SM fermion and $X=B$ a massive electroweak gauge boson ($B=W,Z$). The squared matrix elements $|\mathcal{M}_{{\rm DM}\, X \leftrightarrow {\rm DM}\, X}|^2$ are summed over initial and final spin and polarization states. For details of their implementation, see Ref.~\cite{Binder:2017rgn}. The corresponding expressions are
\begin{flalign} 
|\mathcal{M}_{{\rm S}\, f \leftrightarrow {\rm S}\, f}|^2&= 2 \,c\, \eta_f \, (\lambda_{HS} m_f)^2 \frac{(4 m_f^2-t)}{(t-m_h^2)^2},  \\
|\mathcal{M}_{{\rm S}\, B \leftrightarrow {\rm S}\, B}|^2&= \eta_B \, \lambda_{HS}^2 \frac{(12m_B^4-4m_B^2t+t^2)}{(t-m_h^2)^2}, \\
	|\mathcal{M}_{{\rm V}\, f \leftrightarrow {\rm V}\, f}|^2&= c\, \eta_f \frac{(\lambda_{HV} \, m_f)^2}{2 m_V^4}\frac{(t^2+12 m_V^4-4 \, t \, m_V^2)\,(4\, m_f^2-t)}{(t-m_h^2)^2} , \\
|\mathcal{M}_{{\rm V}\, B \leftrightarrow {\rm V}\, B}|^2&= \eta_B\frac{\lambda_{HV}^2}{4\, m_V^2}\frac{(12 m_B^4+t^2-4\, m_B^2\, t)(12 m_V^4+t^2-4\, m_V^2\, t)}{(t-m_h^2)^2} , \\
	|\mathcal{M}_{{\psi}\, f \leftrightarrow {\psi}\, f}|^2&= 4 \, c \, \eta_f \, (\tilde{\lambda}_{H\psi} m_f)^2\frac{(4\,m_f^2-t)(4 m_\psi^2 \cos{\xi}^2)}{(t-m_h^2)^2} , \\
|\mathcal{M}_{{ \psi}\, B \leftrightarrow {\psi}\, B}|^2&=2\, \eta_B\, \tilde{\lambda}_{H\psi}^2 \frac{(12 m_B^4+t^2-4\, m_B^2 \,t)(4 m_\psi^2 \cos{\xi}^2-t)}{(t-m_h^2)^2}.  
\end{flalign} 
Here $c$ is the colour multiplicity, with $c=3$ for quarks and $c=1$ for leptons. %Since the matrix element is the same for particles and antiparticles, this amounts to 
The contributions from particles and antiparticles are included through
an overall factor $\eta_f=2$ for each fermion species. Similarly, for $B=W$, the two charged states $W^+$ and $W^-$ contribute equally and are accounted for by a multiplicity factor $\eta_W=2$, whereas $\eta_Z=1$ for the neutral boson $B=Z$.
\bibliography{paper}
\bibliographystyle{JHEP-edit}

\end{document}